\documentstyle[twoside,12pt]{article}
\def\bbbr{\rm I\!R}
\def\bbbc{C\!\!\!\!I}
\def\bbbq{Q\!\!\!\!I}
\def\bbbz{Z\!\!\!Z}

\newtheorem{theorem}{Theorem}
\newtheorem{lemma}{Lemma}
\newtheorem{corollary}{Corollary}
\newtheorem{proposition}{Proposition}
\newtheorem{definition}{Definition}
\newtheorem{example}{Example}
\newtheorem{remark}{Remark}
\newtheorem{question}{Question}

\newcommand{\tr}{{\rm Tr\ }}
\newcommand{\coker}{{\rm Coker\ }}
\newcommand{\fix}{{\rm Fix\ }}
\newcommand{\rank}{{\rm Rank\ }}
\newcommand{\spec}{{\rm Spec\ }}
\newcommand{\im}{{\rm Im\ }}
\newcommand{\ind}{{\rm Index\ }}
\newcommand{\mod}{{\rm \; mod\ }}

\date{}

\title{Trace formulas and dynamical zeta functions in the Nielsen theory}
\author{Alexander Fel'shtyn\thanks{Part of this work was conducted during
authors' statement in Sonderforschungsbereich 170 ``Geometrie und Analysis'',
Mathematisches Institut der Georg August Universit\"at zu G\"ottingen.}
\and
 Richard Hill}
\begin{document}
\bibliography{ref}
\bibliographystyle{plain}
\maketitle

\bigskip

 \begin{abstract}
In this paper we prove the trace formulas for the Reidemeister numbers of group
endomorphisms in the following cases:the group is finitely generated and an
endomorphism is eventually commutative; the group is finite ; the group is a
direct sum of a finite group and a finitely generated free abelian group; the
group is finitely generated, nilpotent and torsion free . These results had
previously been known only for the finitely generated free abelian group.As a
consequence, we obtain under the same conditions
on the fundamental group of a compact polyhedron,the trace formulas for the
Reidemeister numbers of a continuous map  and  under suitable conditions the
trace formulas for the Nielsen numbers of a continuous map.The trace formula
for the Reidemeister numbers implies the rationality of the Reidemeister zeta
function. We prove the rationality and functional equation for  the
Reidemeister zeta function of an endomorphisms of finitely generated torsion
free nilpotent group and of a direct sum of a finite group and a finitely
generated free abelian group.We give  a new proof of the rationality of the
Reidemeister zeta function in the case when the group is finitely generated and
the endomorphism is eventually commutative and in the case when group is
finite.We give also another proof for the positivity of the radius of
convergence of the Nielsen zeta function and propose an exast algebraic lower
bound estimation for the radius.We connect the Reidemeister zeta function of an
endom!
 orphism of a direct sum of a finite group and a finitely generated free
abelian group with the Lefschetz zeta function of the unitary dual map,and as
consequence obtain a connection of the Reidemeister zeta function with
Reidemeister torsion.

\end{abstract}

\medskip

\setcounter{section}{-1}

\section{Introduction}
We assume  everywhere $X$ to be a connected, compact
polyhedron and $f:X\rightarrow X$ to be a continuous map. Taking a dynamical
point on view,we consider the iterates of $f$.In the theory of  discrete
dynamical systems the following zeta functions are known: the Artin-Mazur zeta
function
$$
 \zeta_f(z) := \exp\left(\sum_{n=1}^\infty \frac{F(f^n)}{n} z^n \right),
$$
where $F(f^n)$ is the number of isolated fixed points of $f^n$; the Ruelle zeta
function \cite{ruel}
$$
 {\zeta_f}^g(z) := \exp\left(\sum_{n=1}^\infty \frac{ z^n }{n}\sum_{x\in
Fix(f^n)} \prod_{k=0}^{n-1}g(f^k(x)) \right),
$$
where $g:X\to \bbbc $ is a weight function(if $g=1$ we recover $\zeta_f(z) $);
 the Lefschetz zeta function
 $$
 L_f(z) := \exp\left(\sum_{n=1}^\infty \frac{L(f^n)}{n} z^n \right),
$$
where
$$
 L(f^n) := \sum_{k=0}^{\dim X} (-1)^k \tr\Big[f_{*k}^n:H_k(X;\bbbq)\to
H_k(X;\bbbq)\Big]
$$
are the Lefschetz numbers of the iterates of $f$; reduced mod 2 Artin-Mazur and
 Lefschetz zeta functions \cite{fran}; twisted Artin-Mazur and  Lefschetz zeta
functions \cite{fri},which have coefficients in the group rings $\bbbz H$ of an
abelian group $H$.

The above zeta functions are directly analogous to the Hasse-Weil zeta function
of an algebraic manifold over finite fields \cite {weil}.The Lefschetz zeta
function is always rational function of $z$ and
is given by the formula:
$$
 L_f(z) = \prod_{k=0}^{\dim X}
          \det\big(E-f_{*k}.z\big)^{(-1)^{k+1}}.
$$
This immediately follows from trace formula for  the Lefschetz numbers of the
iterates of $f$.The Artin-Mazur zeta function has a positive radius of
convergence for a dence set in the space of smooth maps of a compact smooth
manifold to itself \cite{am}.Manning proved the rationality of the Artin-Mazur
zeta function for diffeomorphisms of a compact smooth manifold satisfying
Smale's  Axiom  A \cite{m}.

The Artin-Mazur zeta function and its modification count periodic points of a
map geometrically, the Lefschetz's type zeta functions do this algebraically
(with weight given by index theory).Another way to count the  periodic points
is given by Nielsen theory.

Let $p:\tilde{X}\rightarrow X$ be the universal covering of $X$
and $\tilde{f}:\tilde{X}\rightarrow \tilde{X}$ a lifting
of $f$, ie. $p\circ\tilde{f}=f\circ p$.
Two liftings $\tilde{f}$ and $\tilde{f}^\prime$ are called
{\it conjugate} if there is a $\gamma\in\Gamma\cong\pi_1(X)$
such that $\tilde{f}^\prime = \gamma\circ\tilde{f}\circ\gamma^{-1}$.
The subset $p(Fix(\tilde{f}))\subset Fix(f)$ is called
{\it the fixed point class of $f$ determined by the lifting class
$[\tilde{f}]$}.
A fixed point class is called $essential$ if its index is nonzero.
 The number of lifting classes of $f$ (and hence the number
of fixed point classes, empty or not) is called the {\it Reidemeister Number}
of $f$,
denoted $R(f)$. This is a positive integer or infinity.
The number of essential fixed point classes is called the $Nielsen$ $number$
of $f$, denoted by $N(f)$.
The Nielsen number is always finite. $R(f)$ and $N(f)$ are homotopy
type invariants.
In the category of compact, connected polyhedra the Nielsen number
of a map is equal to the least number of fixed points
of maps with the same homotopy type as $f$.
Let $G$ be a group and $\phi:G\rightarrow G$ an endomorphism.
Two elements $\alpha,\alpha^\prime\in G$ are said to be
$\phi-conjugate$ iff there exists $\gamma \in G$ such that
$\alpha^\prime=\gamma . \alpha . \phi(\gamma)^{-1}$.
The number of $\phi$-conjugacy classes is called the $Reidemeister$
$number$ of $\phi$, denoted by $R(\phi)$.

If we consider the iterates of $f$ and $\phi$, we may define several zeta
functions connected
with Nielsen fixed point theory (see \cite{f1,f2,f4,f5,fp}).
We assume throughout this article that $R(f^n)<\infty$ and
$R(\phi^n)<\infty$ for all $n>0$.
The Reidemeister zeta functions of $f$ and $\phi$
and the Nielsen zeta function of $f$ are defined
as power series:
\begin{eqnarray*}
 R_\phi(z) & := & \exp\left(\sum_{n=1}^\infty \frac{R(\phi^n)}{n} z^n \right),
\\
 R_f(z) & := & \exp\left(\sum_{n=1}^\infty \frac{R(f^n)}{n} z^n \right), \\
 N_f(z) & := & \exp\left(\sum_{n=1}^\infty \frac{N(f^n)}{n} z^n \right).
\end{eqnarray*}
$R_f(z)$ and $N_f(z)$ are homotopy invariants.
The Nielsen zeta function $N_f(z)$ has a positive radius
of convergence which has a sharp estimate in
terms of the topological entropy of the map $f$ \cite{fp}.In section 4 we
propose another prove of positivity of the radius  and give an exact algebraic
lower estimation for radius using the Reidemeister trace formula for the
generalised Lefschetz numbers.In section 7 we prove the same result for the
radius of convergence of the minimal dynamical zeta function.

  An endomorphism $\phi : G\rightarrow G $ is said to be eventually commutative
if there exists a natural number $n$ such that the subgroup $\phi^n(G) $ is
commutative.
A map $f:X\rightarrow X $ is said to be eventually commutative if the induced
endomorphism on fundamental group is eventually commutative.
We begin the article by proving  in section 1 the trace formula for the
Reidemeister numbers in the following cases:

{\em (I)} $G$ is finitely generated and $\phi$ is eventually commutative.

{\em (II)} $G$ is finite

{\em (III)} $G$ is a direct sum of a finite group and a finitely generated free
abelian group.

{\em (IV)} $G$ is a finitely generated torsion free nilpotent group.

 These result had previously been known only for the finitely generated free
abelian group \cite{f5}.
As a consequence, we obtain in section 2, under the same conditions (I) - (IV)
on the fundamental group of $X$ ,the trace formula for the Reidemeister numbers
of a continuous map  and in section 3 under suitable conditions the trace
formulas for the Nielsen numbers of a continuous map.

 Trace formula for the Reidemeister numbers implies  the rationality of the
Reidemeister zeta function.In section 5 we prove the rationality and functional
equation for  the Reidemeister zeta function of an endomorphisms of finitely
generated torsion free nilpotent group and of a direct sum of a finite group
and a finitely generated free abelian group.We give also  a new proof of the
rationality of $R_{\phi}(z)$ in the case when  $G$ is finitely generated and
$\phi$ is eventually commutative and in the case when $G$ is finite.

 In section 7 we obtain an expression for the Reidemeister torsion
of the mapping torus of the dual map of a endomorphism of a direct sum of a
finite group and a finitely generated free abelian group, in terms of the
Reidemeister zeta function of the endomorphism.The result is obtained by
expressing the Reidemeister zeta function in terms of the Lefschetz zeta
function of the dual map, and then applying the theorem of D. Fried.What this
means is that the Reidemeister torsion counts the fixed point classes of all
iterates of map $f$ i.e. periodic point classes of $f$.These result had
previously been known for the finitely generated abelian groups and finite
groups\cite{fh2}.

In section 5.5 we describe some conjectures on how the Reidemeister
 zeta functions should look in general, largely in terms of the
 growth on the group. In particular we expect that for polynomial growth groups
 some power of the Reidemeister zeta function is a rational function.
\medskip

We would like to thank J.Eichhorn , D.Fried, M.L.Gromov and B.B.Venkov for
valuable conversations and comments.Alexander Fel'shtyn would like to thank
Institut des Hautes Etudes Scientifiques
  for the kind hospitality and support.

\section[Group endomorphisms]{Trace formula for the Reidemeister numbers of a
group endomorphism}

\subsection{Pontryagin Duality}

Let $G$ be a locally compact abelian topological group.
We write $\hat{G}$ for the set of continuous
homomorphisms from $G$ to the circle $U(1)=\{z\in\bbbc : |z|=1\}$.
 This is a group with pointwise multiplication. We call $\hat{G}$
the {\it Pontryagin dual} of $G$. When we equip $\hat{G}$ with
the compact-open topology it becomes a locally compact abelian
topological group. The dual of the dual of $G$ is canonically
isomorphic to $G$.

A continuous endomorphism $f:G\rightarrow G$ gives rise
to a continuous endomorphism $\hat{f}:\hat{G}\rightarrow \hat{G}$ defined by
$$
 \hat{f}(\chi) := \chi\circ f.
$$
There is a 1-1 correspondence between the closed subgroups $H$ of $G$
and the quotient groups $\hat{G}/H^*$ of $\hat{G}$ for which
$H^*$ is closed in $\hat{G}$.
This correspondence is given by the following:
 $$  H \leftrightarrow \hat{G}/H^*, $$
 $$
 H^* := \{\chi \in \hat{G} \mid H\subset\ker\chi\}.
 $$
Under this correspondence, $\hat{G}/H^*$ is canonically
isomorphic to the Pontryagin dual of $H$.
If we identify $G$ canonically with the dual of $\hat{G}$
then we have $H^{**}=H$.
 If $G$ is a finitely generated  abelian group then a homomorphism
$\chi:G\rightarrow U(1)$ is completely
determined by its values on a basis of $G$, and
these values may be chosen arbitrarily. The dual of $G$
is thus a torus whose dimension is equal to the rank of $G$.

If $G=\bbbz/n \bbbz$ then the elements of $\hat{G}$
are of the form
 $$ x \rightarrow e^{\frac{2\pi i y x}{n}} $$
with $y\in\{1,2,\ldots,n\}$.
A cyclic group is therefore (uncanonically) isomorphic to
itself.

The dual of $G_1\oplus G_2$ is canonically isomorphic
to $\hat{G}_1 \oplus\hat{G}_2$. From this we see that
any finite abelian group is (non-canonically) isomorphic to
its own Pontryagin dual group, and that the dual of any finitely generated
discrete abelian group
 is the direct sum of a Torus and a
finite group.

Proofs of all these statements may be found, for example in
 \cite{ru}.
We shall require the following statement:

\begin{proposition}
Let $\phi:G\to G$ be an endomorphism of an abelian group $G$.
Then the kernel $\ker\left[\hat{\phi}:\hat{G}\to\hat{G}\right]$
is canonically isomorphic to the
Pontryagin dual of $\coker\phi$.
\end{proposition}

{\sc Proof}
We construct the isomorphism explicitly.
Let $\chi$ be in the dual of $\coker(\phi:G\to G)$.
In that case $\chi$ is a homomorphism
 $$  \chi : G / \im(\phi) \longrightarrow U(1).$$
There is therefore an induced map
 $$  \overline{\chi} : G \longrightarrow U(1) $$
which is trivial on $\im(\phi)$.
This means that $\overline{\chi}\circ\phi$ is trivial, or
in other words $\hat{\phi}(\overline{\chi})$ is the identity element of
$\hat{G}$.
We therefore have $\overline{\chi}\in\ker(\hat{\phi})$.
If on the other hand we begin with $\overline{\chi}\in\ker(\hat{\phi})$,
then it follows that $\chi$ is trivial on $\im\phi$, and so
$\overline{\chi}$ induces a homomorphism
 $$ \chi : G / \im(\phi) \longrightarrow U(1)$$
and $\chi$ is then in the dual of $\coker\phi$.
The correspondence $\chi\leftrightarrow\overline{\chi}$ is
clearly a bijection.

\subsection{Eventually commutative endomorphisms}

\subsubsection{Trace formula for the Reidemeister numbers of eventually
commutative endomorphisms.}

An endomorphism $\phi:G\rightarrow G$ is said to be eventually
commutative if there exists a natural number $n$ such that the
subgroup $\phi^n(G)$ is commutative.
If $\phi$ is an endomorphism of an abelian group $G$ then $x$ and $y$ are
$\phi-conjugate$ iff $ x-y =\phi(g) -g $ for some $g\in G$.Therefore $R(\phi)$
is the number of cosets of the image of the endomorphism $ (\phi -
1):G\rightarrow G$.Then $$  R(\phi) = \#\coker(1-\phi) .$$
We are now ready to compare the Reidemeister number  of an
endomorphism $\phi$ with the Reidemeister number
 of $H_1(\phi):H_1(G)\rightarrow H_1(G)$,
where $H_1 = H_1^{Gp}$ is the first integral homology functor
from groups to abelian groups.

\begin{lemma}[\cite{j}]
If $\phi:G\rightarrow G$ is eventually commutative,
then
$$
 R(\phi) = R({H_1(\phi)}= \#\coker(1- H_1(\phi))
$$
\end{lemma}

This means that to find out about the Reidemeister numbers
of eventually commutative endomorphisms, it is sufficient to
study the Reidemeister numbers of endomorphisms of abelian groups.
For the rest of this section G will be a finitely generated
abelian group.

\begin{lemma}[\cite{f5}]
Let $\phi:\bbbz^k\rightarrow\bbbz^k$ be a group endomorphism.
Then we have
 \begin{equation}
R(\phi) = (-1)^{r+p} \sum_{i=0}^k (-1)^i \tr (\Lambda^i\phi).
\end{equation}
where $p$ the number of $\mu\in\spec\phi$ such that
$\mu <-1$, and $r$ the number of real eigenvalues of $\phi$ whose
absolute value is $>1$.
$\Lambda^i$ denotes the exterior power.
\end{lemma}

{\sc Proof}
Since $\bbbz^k$ is abelian, we have as before,
 $$  R(\phi) = \#\coker(1-\phi) .$$
On the other hand we have
 $$ \#\coker(1-\phi) = \mid\det(1-\phi)\mid ,$$
and hence $R(\phi)=(-1)^{r+p}\det(1-\phi)$ ( complex eigenvalues contribute
nothing to the sign$\det(1-\phi)$ since they come in conjugate pairs and $
(1-\lambda)(1-\bar \lambda)=\mid 1-\lambda \mid^2 > 0$).
It is well known from linear algebra that
$\det(1-\phi)=\sum_{i=0}^k (-1)^i \tr (\Lambda^i\phi)$.
{}From this we have the  trace formula for Reidemeister number.

 Now let  $\phi$ be an endomorphism of finite abelian group $G$.Let $V$ be the
complex vector space of complex valued functions on the group $G$.The map
$\phi$ induces a linear map $A:V\rightarrow V$ defined by
$$
A(f):=f\circ\phi.
$$

\begin{lemma}
Let $\phi:G\rightarrow G$ be an endomorphism
of a finite abelian group $G$. Then we have
\begin{equation}
R(\phi) = TrA
\end{equation}
\end{lemma}
We give two proofs of this lemma in this article.
The first proof is given here and the second
proof is a special case of the proof of theorem 4

{\sc Proof}
We shall calculate the trace of $A$ in two ways.The characteristic functions of
the elements of $G$ form a basis of $V$,and are mapped to one another by
$A$(the  map need not be a bijection).Therefore the trace of $A$ is the number
of
elements of this basis which are fixed by $A$.On the other hand, since $G$ is
abelian, we  have,
\begin{eqnarray*}
 R(\phi)
 & = &
 \#\coker(1-\phi) \\
 & = &
 \# G / \#\im(1-\phi) \\
 & = &
 \# G / \#(G/\ker(1-\phi)) \\
 & = &
 \# G / (\#G /\#\ker(1-\phi)) \\
 & = &
 \#\ker(1-\phi) \\
 & = &
 \#\fix(\phi)
\end{eqnarray*}
We therefore have  $ R(\phi)=\#\fix(\phi)=TrA $ .

For a finitely generared abelian group $G$ we define the finite subgroup
$TorsG$ to be the subgroup of
torsion elements of $G$. We denote the quotient $G^\infty:=G/TorsG$.
The group $G^\infty$ is torsion free.
Since the image of any torsion element by a homomorphism must be a torsion
element, the endomorphism $\phi:G\to G$ induces endomorphisms
 $$
 \phi^{tor}:TorsG\longrightarrow TorsG,\;\;\;\;
 \phi^\infty:G^\infty\longrightarrow G^\infty.
 $$

As above, the map $\phi^{tor}$ induces a linear map $A:V\rightarrow V$, where
$V$ be the complex vector space of complex valued  functions on the group
$TorsG$.

\begin{theorem}
If $G$ is a finitely generated abelian group and $\phi$ an
endomorphism of $G$ .Then we have
\begin{equation}
R(\phi) = (-1)^{r+p} \sum_{i=0}^k (-1)^i \tr (\Lambda^i\phi^\infty\otimes A).
\end{equation}
where $k$ is $rgG^\infty$, $p$ the number of $\mu\in\spec\phi^\infty$ such that
$\mu <-1$, and $r$ the number of real eigenvalues of $\phi^\infty$ whose
absolute value is $>1$.
\end{theorem}

{\sc Proof}
By proposition 1, the cokernel of $(1-\phi):G\rightarrow G$ is the Pontrjagin
dual of the
kernel of the dual map $\widehat{(1-\phi)}:\hat{G}\rightarrow \hat{G}$.
Since $\coker (1-\phi)$ is finite, we have
  $$
 \#\coker (1-\phi) = \#\ker \widehat{(1-\phi)} .
 $$
The map $\widehat{1-\phi}$ is equal to $\hat{1}-\hat{\phi}$.
Its kernel is thus the set of fixed points of the
map $\hat{\phi}:\hat{G}\rightarrow \hat{G}$.
We therefore have
\begin{equation}
 R(\phi) = \#\fix\left(\hat{\phi}:\hat{G}\rightarrow \hat{G}\right)
\end{equation}
The dual group of $G^\infty$ is a torus whose dimension is the
rank of $G$. This is canonically a closed subgroup of $\hat{G}$.
We shall denote it $\hat{G}_0$.
The quotient $\hat{G}/\hat{G}_0$ is canonically isomorphic to the
dual of $TorsG$. It is therefore finite.
{}From this we know that $\hat{G}$ is a union of finitely many disjoint
tori. We shall call these tori $\hat{G}_0,\ldots,\hat{G}_t$.

We shall call a torus $\hat{G}_i$ periodic if there is an iteration
$\hat{\phi}^s$ such that $\hat{\phi}^s(\hat{G}_i)\subset\hat{G}_i$.
If this is the case, then the map $\hat{\phi}^s:\hat{G}_i\rightarrow\hat{G}_i$
is a translation of the map $\hat{\phi}^s:\hat{G}_0\rightarrow\hat{G}_0$ and
has
 the same number of fixed points as this map.
If $\hat{\phi}^s(\hat{G}_i)\not\subset\hat{G}_i$ then $\hat{\phi}^s$ has
no fixed points in $\hat{G}_i$.
{}From this we see
 $$
 \#\fix\left(\hat{\phi}:\hat{G}\rightarrow \hat{G}\right)
 =
 \#\fix\left(\hat{\phi}:\hat{G}_0\rightarrow \hat{G}_0\right)
 \times
 \#\{\hat{G}_i \mid \hat{\phi}(\hat{G}_i)\subset\hat{G}_i\}.
 $$
We now rephrase this
\begin{eqnarray*}
\lefteqn{ \#\fix\left(\hat{\phi}:\hat{G}\rightarrow \hat{G}\right)}\\
& = &
 \#\fix\left(
 \widehat{\phi^\infty}:\hat{G}_0\rightarrow \hat{G}_0
 \right)
 \times
 \#\fix\left(
 \widehat{\phi^{tor}}:\hat{G}/(\hat{G}_0) \rightarrow \hat{G}/(\hat{G}_0)
 \right).
\end{eqnarray*}
 From this we have product formula for Reidemeister numbers
 $$ R(\phi) = R(\phi^\infty) . R(\phi^{tors}). $$
The trace formula for $R(\phi)$ follow from the previous two lemmas and formula
$$
Tr(\Lambda^i\phi^\infty).Tr(A)=Tr(\Lambda^i\phi^\infty\otimes A).
$$

In the paper \cite{fh2} we have connected  Reidemeister number of endomorphism
$\phi$ with Lefschetz number of the dual map.From this we have following trace
formula

\begin{theorem}[\cite{fh2}]
Let $\phi:G\to G$ be an endomorphism of a finitely
generated abelian group.
Then
\begin{equation}
  R(\phi) = \mid L(\hat{\phi}) \mid = (-1)^{r+p} \sum_{k=0}^{\dim\hat{G} }
(-1)^k \tr\Big[\hat{\phi}_{*k}:H_k(\hat{G};\bbbq)\to H_k(\hat{G};\bbbq)\Big]
\end{equation}
where $\hat{\phi}$ is the continuous endomorphism of $\hat{G}$
 defined in \S 2.2 and $L(\hat{\phi})$ is the Lefschetz number
of $\hat{\phi}$ thought of as a self-map of the topological space $\hat{G}$
and $r$ and $p$ are the constants described in theorem 1.
If $G$ is finite then this reduces to
$$  R(\phi)=L(\hat{\phi})=  \tr\Big[\hat{\phi}_{*0}:H_0(\hat{G};\bbbq)\to
H_0(\hat{G};\bbbq)\Big].           $$

\end{theorem}

\subsection{Endomorphisms of finite groups}

In this section we consider finite non-abelian groups.
We shall write the group law multiplicatively.
We generalize our results on endomorphisms
of finite abelian groups to endomorphisms of finite non-abelian groups.
We shall write $\{g\}$ for the
$\phi$-conjugacy class of an element $g\in G$. We shall write
$<g>$ for the ordinary conjugacy class of $g$ in $G$.
We first note that if $\phi$ is an endomorphism of a group $G$
then $\phi$ maps conjugate elements to conjugate elements. It
therefore induces an endomorphism of the set of conjugacy classes of elements
of $G$.
If $G$ is abelian then a conjugacy class of element consists of
a single element.
The following is thus an extension of lemma 3:

\begin{theorem}[\cite{fh1}]
Let $G$ be a finite group and let $\phi:G\to G$ be an endomorphism.
Then $R(\phi)$ is the number of ordinary conjugacy classes $<x>$ in
$G$ such that
 $$
 < \phi(x)>=<x>.
 $$
\end{theorem}
{\sc Proof}
{}From the definition of the Reidemeister number we have,
\begin{eqnarray*}
 R(\phi)
 & = &
 \sum_{\{g\}} 1
\end{eqnarray*}
where $\{g\}$ runs through the
set of $\phi$-conjugacy classes in $G$.
This gives us immediately
\begin{eqnarray*}
 R(\phi)
 & = &
 \sum_{\{g\}} \sum_{x\in\{g\}} \frac{1}{\#\{g\}}  \\
 & = &
 \sum_{\{g\}} \sum_{x\in\{g\}} \frac{1}{\#\{x\}}  \\
 & = &
 \sum_{x\in G} \frac{1}{\#\{x\}}.
\end{eqnarray*}
We now calculate for any $x\in G$ the order of $\{x\}$.
The class $\{x\}$ is the orbit of $x$ under the $G$-action

 $$  (g , x)\longmapsto g x \phi(g)^{-1}  .$$
We verifty that this is actually a $G$-action:
\begin{eqnarray*}
 (id,x)
 & \longmapsto & id.x. \phi(id)^{-1} \\
 & = & x,  \\
 (g_1 g_2,x)
 & \longmapsto & g_1 g_2.x. \phi(g_1 g_2)^{-1} \\
 & = & g_1 g_2.x. (\phi(g_1) \phi(g_2))^{-1} \\
 & = & g_1 g_2.x. \phi(g_2)^{-1} \phi(g_1)^{-1} \\
 & = & g_1 (g_2.x. \phi(g_2)^{-1}) \phi(g_1)^{-1}.
\end{eqnarray*}
We therefore have from the orbit-stabilizer theorem,
 $$ \#\{x\} = \frac{\#G}{\#\{g\in G\mid g x \phi(g)^{-1} = x \} }.$$
The condition $g x \phi(g)^{-1} = x$
is equivalent to
\begin{eqnarray*}
x^{-1} g x \phi(g)^{-1} = 1
& \Leftrightarrow &
x^{-1} g x = \phi(g)
\end{eqnarray*}
We therefore have
$$
 R(\phi)= \frac{1}{\#G} \sum_{x\in G} \#\{ g\in G\mid x^{-1}gx=\phi(g)\}.
$$
Changing the summation over $x$ to summation over $g$, we have:
$$
 R(\phi)= \frac{1}{\#G} \sum_{g\in G} \#\{ x\in G\mid x^{-1}gx=\phi(g)\}.
$$
If $<\phi(g)>\not=<g>$ then there are no elements $x$ such that
$x^{-1}gx=\phi(g)$.
We therefore have:
$$
 R(\phi)
 =
 \frac{1}{\#G}
 \sum_{g\in G \ {\rm such \ that} \atop <\phi(g)>=<g>}
 \#\{ x\in G\mid x^{-1}gx=\phi(g)\}.
$$
The elements $x$ such that $x^{-1}gx=\phi(g)$
form a coset of the subgroup satisfying $x^{-1}gx=g$.
This subgroup is the centralizer of $g$ in $G$ which we write $C(g)$.
With this notation we have,
\begin{eqnarray*}
 R(\phi)
 & = &
  \frac{1}{\#G}
 \sum_{g\in G \ {\rm such \ that} \atop <\phi(g)>=<g>}
 \#C(g) \\
 & = &
 \frac{1}{\#G}
 \sum_{<g>\subset G \ {\rm such \ that} \atop <\phi(g)>=<g>}
 \#<g>. \#C(g).
\end{eqnarray*}
The last identity follows because $C(h^{-1}gh)=h^{-1}C(g)h$.
{}From the orbit stabilizer theorem, we know that
$\#<g>. \#C(g)=\#G$.
We therefore have
$$
 R(\phi) = \#\{<g>\subset G \mid <\phi(g)>=<g>\}.
$$
\vspace{0.2cm}
Let $W$ be the complex vector space of complex valued class functions on the
group $G$.A class function is a function which takes the same value on every
element of a usual congruency class.The map $\phi$ induces a linear map
$B:W\rightarrow W$ defined by
$$
B(f):=f\circ\phi.
$$

\begin{theorem}
Let $\phi:G\rightarrow G$ be an endomorphism
of a finite  group $G$. Then we have
\begin{equation}
R(\phi) = TrB
\end{equation}
\end{theorem}

{\sc Proof}
We shall calculate the trace of $B$ in two ways.The characteristic functions of
the congruency classes in $G$ form a basis of $W$,and are mapped to one another
by $B$(the map need not be a bijection).Therefore the trace of $B$ is the
number of elements of this basis which are fixed by $B$.By theorem 3,this is
equal to the Reidemeister number of $\phi$.

\subsection{Endomorphisms of the direct sum of a free abelian and a finite
group}

In this section let $F$ be a finite group and $r$ a natural number.The group
$G$
will be $G=\bbbz^k\oplus F$.The torsion elements of $G$ are exastly the
elements of the finite, normal subgroup $F$.For this reason we have
$\phi(F)\subset F$.Let $\phi^{tors}: F\to F $ be the restriction of $\phi$ to
$F$, and let $\phi^{\infty}: G/F\to G/F $ be the induced map on the quotient
group.
\begin{lemma}\cite{hi}
In the notation described above
$$
R(\phi)=R(\phi^{\infty}).R(\phi^{tors})
$$
\end{lemma}

Let $W$ be the complex vector space of complex valued class functions on the
group $F$.The map $\phi$ induces a linear map $B:W\rightarrow W$ defined as
above.

\begin{theorem}
If $G$ is the direct sum of a free abelian and a finite group and $\phi$ an
 endomorphism of $G$ .Then we have
\begin{equation}
R(\phi) = (-1)^{r+p} \sum_{i=0}^k (-1)^i \tr (\Lambda^i\phi^\infty\otimes B).
\end{equation}
where $k$ is $rg(G/F)$, $p$ the number of $\mu\in\spec\phi^\infty$ such that
$\mu <-1$, and $r$ the number of real eigenvalues of $\phi^\infty$ whose
absolute value is $>1$.
\end{theorem}

{\sc Proof}
Theorem follows from lemmas 2 and 4 , theorem 4 and formula
$$
Tr(\Lambda^i\phi^\infty).Tr(B)=Tr(\Lambda^i\phi^\infty\otimes B).
$$

\subsection{Endomorphisms of nilpotent groups}

In this section we consider finitely generated torsion free nilpotent group
$\Gamma$.It is well known \cite{mal} that such group $\Gamma$ is a uniform
discrete subgroup of a simply connected nilpotent Lie group $G$ (uniform means
that the coset space $G/ \Gamma$ is compact).The coset space $M=G/ \Gamma$ is
called a nilmanifold.Since $\Gamma=\pi_1(M)$ and $M$  is a $K(\Gamma,
1)$, every endomorphism $\phi:\Gamma \to \Gamma $ can be realized by a selfmap
$f:M\to M$ such that $f_*=\phi$ and thus $R(f)=R(\phi)$.Any endomorphism
$\phi:\Gamma \to \Gamma $ can be uniquely extended to an endomorphism  $F: G\to
G$.Let $\tilde F:\tilde G\to \tilde G $ be the corresponding Lie algebra
endomorphism induced from $F$.

\begin{theorem}
If $G$ is a finitely generated torsion free nilpotent group and $\phi$ an
endomorphism of $G$ .Then
\begin{equation}
R(\phi) = (-1)^{r+p} \sum_{i=0}^m (-1)^i \tr \Lambda^i\tilde F,
\end{equation}

where $m$ is $rg\Gamma = \dim M$, $p$ the number of $\mu\in\spec\tilde F$ such
that
$\mu <-1$, and $r$ the number of real eigenvalues of $\tilde F$ whose
absolute value is $>1$.
\end{theorem}

{\sc Proof:}
Let $f:M\to M$ be a map realizing $\phi$ on a compact nilmanifold $M$ of
dimension $m$.We suppose in this article that the Reidemeister number
$R(f)=R(\phi)$ is finite.The finiteness of $R(f)$ implies the nonvanishing of
the Lefschetz number $L(f)$ \cite{fhw}.A strengthened version of Anosov's
theorem \cite{a} is proven in
\cite{n} which states, in particular, that if
$L(f)\ne 0$ than $N(f)=|L(f)|=R(f)$.It is well known that $L(f)=\det (\tilde F
-1)$ \cite{a}.From this we have
$$
R(\phi)=R(f)=|L(f)|=|\det (1- \tilde F)|=(-1)^{r+p}\det (1- \tilde F)=
(-1)^{r+p} \sum_{i=0}^m (-1)^i \tr \Lambda^i\tilde F.
$$

\subsection{The trace formulas and group extensions.}

Suppose we are given a commutative diagram
\begin{equation}
\begin{array}{lcl}
  G             &  \stackrel{\phi}{\longrightarrow}      &  G  \\
  \downarrow p  &                                        &  \downarrow p\\
  \overline{G}  &  \stackrel{\overline{\phi}}{\longrightarrow}      &
\overline{G}
\end{array}
\end{equation}
of groups and homomorphisms.
In addition let the sequence
\begin{equation}
  0 \rightarrow H \rightarrow G \stackrel{p}{\rightarrow} \overline{G}
\rightarrow 0
\end{equation}
be exact.
Then $\phi$ restricts to an endomorphism
$\phi\mid_H : H \rightarrow H$.

\begin{definition}
The short exact sequence (16) of groups
is said to have a {\it normal splitting} if there is a section
$\sigma:\overline{G}\rightarrow G$ of $p$
such that $\im \sigma = \sigma(\overline{G})$ is a normal subgroup of $G$.
An endomorphism $\phi:G\rightarrow G$ is said to
{\it preserve} this normal splitting if $\phi$ induces a
morphism of (16) with $\phi(\sigma(\overline{G}))\subset\sigma(\overline{G})$.
\end{definition}

In this section we study the relation between the traces formulas of type (7)
for
Reidemeister numbers
$R(\phi)$, $R({\overline{\phi}})$ and $R({\phi\mid_H})$.

\begin{theorem}
 Let the sequence (14)
have a normal splitting which is preserved by $\phi:G\rightarrow G$.
If we have a trace formulas of the type (7) for $R({\overline{\phi}})$ and
$R({\phi\mid_H})$
then we have a trace formula of the same type for $ R(\phi)$.

\end{theorem}

{\sc Proof}
{}From the assumptions of the theorem it follows that
$$  R(\phi)
  = R(\overline{\phi}) . R(\phi\mid_H)
  \;\;\;\;\; {\rm (see\ \cite{h}).}  $$
Trace formula for $ R(\phi)$ now follows from linear algebra formula:
$$
TrA.TrB=TrA\otimes B.
$$

\subsubsection{Direct Sums}
If $G=G_1\oplus G_2$ is a direct sum and if $\phi(G_i)\subset G_i$
 for $i=1,2$ then it has been shown (see \cite{h})
 that $R(\phi)=R(\phi_1)\cdot R(\phi_2)$ where $\phi_i$
 is the restriction of $\phi$ to $G_i$.So if we have the trace formula of type
(7) for $R(\phi_1)$ and $ R(\phi_2$) then  we have the trace formula for $
R(\phi)$.

\section[Continuous maps]{ Trace formulas for Reidemeisters numbers of a
continuous map.}

Let $f:X\rightarrow X$ be given, and let a
specific lifting $\tilde{f}:\tilde{X}\rightarrow\tilde{X}$ be chosen
as reference.
Let $\Gamma$ be the group of
 covering translations of $\tilde{X}$ over $X$.
Then every lifting of $f$ can be written uniquely
as $\gamma\circ \tilde{f}$, with $\gamma\in\Gamma$.
So elements of $\Gamma$ serve as coordinates of
liftings with respect to the reference $\tilde{f}$.
Now for every $\gamma\in\Gamma$ the composition $\tilde{f}\circ\gamma$
is a lifting of $f$ so there is a unique $\gamma^\prime\in\Gamma$
such that $\gamma^\prime\circ\tilde{f}=\tilde{f}\circ\gamma$.
This correspondence $\gamma\rightarrow\gamma^\prime$ is determined by
the reference $\tilde{f}$, and is obviously a homomorphism.

\begin{definition}
The endomorphism $\tilde{f}_*:\Gamma\rightarrow\Gamma$ determined
by the lifting $\tilde{f}$ of $f$ is defined by
$$
  \tilde{f}_*(\gamma)\circ\tilde{f} = \tilde{f}\circ\gamma.
$$
\end{definition}

It is well known that $\Gamma\cong\pi_1(X)$.
We shall identify $\pi=\pi_1(X,x_0)$ and $\Gamma$ in the following way.
Pick base points $x_0\in X$ and $\tilde{x}_0\in p^{-1}(x_0)\subset \tilde{X}$
once and for all.
 Now points of $\tilde{X}$ are in 1-1 correspondence with homotopy classes of
paths
in $X$ which start at $x_0$:
for $\tilde{x}\in\tilde{X}$ take any path in $\tilde{X}$ from $\tilde{x}_0$ to
$\tilde{x}$ and project it onto $X$;
conversely for a path $c$ starting at $x_0$, lift it to a path in $\tilde{X}$
which starts at $\tilde{x}_0$, and then take its endpoint.
In this way, we identify a point of $\tilde{X}$ with
a path class $<c>$ in $X$ starting from $x_0$. Under this identification,
$\tilde{x}_0=<e>$ is the unit element in $\pi_1(X,x_0)$.
The action of the loop class $\alpha = <a>\in\pi_1(X,x_0)$ on $\tilde{X}$
is then given by
$$
\alpha = <a> : <c>\rightarrow \alpha . c = <a.c>.
$$
Now we have the following relationship between $\tilde{f}_*:\pi\rightarrow\pi$
and
$$
f_*  :  \pi_1(X,x_0) \longrightarrow \pi_1(X,f(x_0)).
$$

\begin{lemma}
Suppose $\tilde{f}(\tilde{x}_0) = <w>$.
Then the following diagram commutes:
$$
\begin{array}{ccc}
  \pi_1(X,x_0)  &  \stackrel{f_*}{\longrightarrow}  &  \pi_1(X,f(x_0))  \\
                &  \tilde{f}_* \searrow \;\; &  \downarrow w_*   \\
                &                           &  \pi_1(X,x_0)
\end{array}
$$
where $w_*$ is isomorphism induced by the path $w$.
\end{lemma}
In other words, for every  $\alpha = <a>\in\pi_1(X,x_0)$ , we have
$$
\tilde{f}_*(<a>)=<w(f \circ a)w^{-1}>
$$
\begin{remark}
In particular, if $ x_0 \in p(Fix( \tilde f)$ and $ \tilde x_0 \in Fix( \tilde
f)$,then $ \tilde{f}_*=f_*$.
\end{remark}

\begin{lemma}
Lifting classes of $f$ are in 1-1 correspondence with
$\tilde{f}_*$-conjugacy classes in $\pi$,
the lifting class $[\gamma\circ\tilde{f}]$ corresponding
to the $\tilde{f}_*$-conjugacy class of $\gamma$.
We therefore have $R(f) = R(\tilde{f}_*)$.
\end{lemma}

We shall say that the fixed point class $p(Fix(\gamma\circ\tilde{f}))$,
which is labeled with the lifting class $[\gamma\circ\tilde{f}]$,
{\it corresponds} to the $\tilde{f}_*$-conjugacy class of $\gamma$.
Thus $\tilde{f}_*$-conjugacy classes in $\pi$ serve as coordinates for fixed
point classes of $f$, once a reference lifting $\tilde{f}$ is chosen.

Let us consider a homomorphisms from $\pi$ sending
an $\tilde{f}_*$-conjugacy class to one element:

\begin{lemma}[\cite{j}]
The composition $\eta\circ\theta$,
$$
 \pi = \pi_1(X,x_0) \stackrel{\theta}{\longrightarrow} H_1(X)
 \stackrel{\eta}{\longrightarrow}
 \coker\left[H_1(X) \stackrel{1-f_{1*}}{\longrightarrow} H_1(X)\right] ,
$$
where $\theta$ is abelianization and $\eta$ is the natural projection,
sends every $\tilde{f}_*$-conjugacy class to a single element.
Moreover, any group homomorphism $\zeta:\pi\rightarrow G$ which
sends every $\tilde{f}_*$-conjugacy class to a single element,
factors through $\eta\circ\theta$.
\end{lemma}

\begin{definition}
A map $f:X\rightarrow X$ is said to be {\it eventually commutative}
if there exists an natural number $n$ such that
$f^n_*(\pi_1(X,x_0))\;\; (\subset \pi_1(X,f^n(x_0)))$ is commutative.
\end{definition}

By means of Lemma 5, it is easily seen that $f$ is eventually
commutative iff $\tilde{f}_*$ is eventually commutative (see \cite{j}).

Now using lemma 6 we may apply all theorems of \S1 to the Reidemeister
numbers of continuous maps.

\subsection{Trace formulas and Serre bundles.}
For example, let us consider topological counterpart of theorem 7.

Let $p:E\rightarrow B$ be a Serre bundle in which $E$, $B$ and
every fibre are connected, compact polyhedra and $F_b=p^{-1}(b)$
is a fibre over $b\in B$.
A Serre bundle $p:E\rightarrow B$ is said to be
{\it (homotopically) orientable} if for any two paths $w$, $w^\prime$ in $B$
with the same endpoints $w(0)= w^\prime(0)$ and $w(1)= w^\prime(1)$,
the fibre translations
$\tau_w , \tau_{w^\prime} : F_{w(0)}\rightarrow F_{w(1)}$ are homotopic.
A map $f:E\rightarrow E$ is called a {\it fibre map} if there
is an induced map $\bar{f}:B\rightarrow B$ such that
$p\circ f = \bar{f}\circ p$.
Let $p:E\rightarrow B$ be an orientable Serre bundle and
let $f:E\rightarrow E$ be a fibre map.
Then for any two fixed points $b,b^\prime$ of $\bar{f}:B\rightarrow B$
the maps $f_b=f\mid_{F_b}$ and $f_{b^\prime} = f\mid_{F_{b^\prime}}$
have the same homotopy type;
 hence they have the same Reidemeister numbers
$R(f_b) = R(f_{b^\prime})$ \cite{j}.

\begin{theorem}
Suppose that $f:E \rightarrow E$ admits a Fadell splitting
in the sense that for some $e$ in $\fix f$ and $b=p(e)$ the
following conditions are satisfied:
\begin{enumerate}
\item
the sequence
$$  0 \longrightarrow \pi_1(F_b,e)
      \stackrel{i_*}{\longrightarrow} \pi_1(E,e)
      \stackrel{p_*}{\longrightarrow} \pi_1(B,e)
      \longrightarrow 0 $$
is exact,

\item
$p_*$ admits a right inverse (section) $\sigma$ such that $\im\sigma$
is a normal subgroup of $\pi_1(E,e)$ and $f_*(\im\sigma)\subset\im\sigma$.

\end{enumerate}
If we have a trace formulas of the type (7) for $R({\bar{f}})$ and $R({f_b})$
then we have a trace formula of the same type for $ R(f)$.
\end{theorem}

\section{Trace formulas for the Nielsen numbers}

\subsection{The Jiang subgroup and trace formula}

{}From the homotopy invariance theorem (see \cite{j})
it follows that if a homotopy $\{h_t\}:f\cong g:X\rightarrow X$
lifts to a homotopy
 $\{\tilde{h}_t\}:\tilde{f}\cong \tilde{g}:\tilde{X}\rightarrow \tilde{X}$,
 then we have
 ${\rm Index}(f,p(Fix\ \tilde{f})) = {\rm Index}(g,p(Fix\ \tilde{g}))$.
Suppose $\{h_t\}$ is a cyclic homotopy $\{h_t\}:f\cong f$;
then this lifts to a homotopy from a given lifting $\tilde{f}$ to
another lifting $\tilde{f}^\prime = \alpha\circ\tilde{f}$, and we have

 $$
 {\rm Index}(f,p(Fix\ \tilde{f})) = {\rm Index}(f,p(Fix\
\alpha\circ\tilde{f})).
 $$
In other words, a cyclic homotopy induces a permutation of lifting classes
(and hence of fixed point classes);
those in the same orbit of this permutation have the same index.
This idea is applied to the computation of $N(f)$.

\begin{definition}
The {\it trace subgroup of cyclic homotopies} (the {\it Jiang subgroup})
$I(\tilde{f})\subset\pi$ is defined by
$$
 I(\tilde{f})
 =
 \left\{ \alpha\in\pi \ \Bigg|
 \begin{array}{l}
    {\rm there\ exists\ a\ cyclic\ homotopy\ } \\
    \{h_t\}:f\cong f
    {\rm which\ lifts\ to\ } \\
    \{\tilde{h}_t\}:
    \tilde{f}\cong\alpha\circ\tilde{f}
  \end{array}
 \right\}
$$
(see \cite{j}).
\end{definition}

Let $Z(G)$ denote the centre of a group $G$, and let $Z(K,G)$
denote the centralizer of the subgroup $K\subset G$.
The Jiang subgroup has the following properties:
$1. I(\tilde{f})\subset Z(\tilde{f}_*(\pi),\pi);\\
 2.   I(id_{\tilde{X}}) \subset Z(\pi) ;3.   I(\tilde{g}) \subset
I(\tilde{g}\circ\tilde{f});  4.   \tilde{g}_*(I(\tilde{f}))\subset
I(\tilde{g}\circ\tilde{f});  5.   I(id_{\tilde{X}}) \subset I(\tilde{f}).$

The class of path-connected spaces $X$ satisfying the condition
$I(id_{\tilde{X}})=\pi=\pi_1(X,x_0)$ is closed under homotopy equivalence
and the topological product operation, and contains the simply connected
spaces, generalized lens spaces, $H$-spaces and homogeneous spaces of the form
$G/G_0$ where $G$ is a topological group and $G_0$ a subgroup which is a
connected, compact Lie group (for the proofs see \cite{j}).

{}From theorem 1 and results of Jiang \cite{j} it follows:

\begin{theorem}
Suppose that there is an integer $m$ such that $\tilde{f}_*^m(\pi)\subset
I(\tilde{f}^m)$ and $L(f)\not=0$.Then
\begin{equation}
N(f)= R(f)= (-1)^{r+p} \sum_{i=0}^k (-1)^i tr (\Lambda^i {f_{1*}}^\infty
\otimes A).
\end{equation}
where $k$ is $rgH_1(X,\bbbz)^\infty$,$A$ is linear map on the complex vector
space of complex valued functions on the group $TorsH_1(X,\bbbz)$, $p$ the
number of $\mu\in\spec f_{1*}^\infty$ such that
$\mu <-1$, and $r$ the number of real eigenvalues of $f_{1*}^\infty$ whose
absolute value is $>1$.
\end{theorem}
{\sc Proof}
We have $\tilde{f}_*(\pi)\subset I(\tilde{f})$(see \cite{j}) . For any
$\alpha\in\pi$,
$p(\fix \alpha\circ\tilde{f})=p(\fix \tilde{f}_*(\alpha)\circ\tilde{f})$
by the fact (see \cite{j}) that $\alpha$ and $\tilde{f}_*(\alpha)$ are in the
same $\tilde{f}_*$-conjugacy class.

Since $\tilde{f}_*(\pi)\subset I(\tilde{f})$, there is
a homotopy $\{h_t\}:f\cong f$ which lifts to
$\{\tilde{h}_t\}:\tilde{f}\cong \tilde{f}_*(\alpha)\circ\tilde{f}$.
Hence $\ind(f,p(\fix \tilde{f}))=\ind(f,p(\fix \alpha\circ\tilde{f}))$.
Since $\alpha\in\pi$ is arbitrary, any two fixed point classes of
$f$ have the same index.
It immediately follows that
$L(f)=0$ implies $N(f)=0$ and $L(f)\not=0$ implies
$N(f)=R(f)$.
By property 1, $\tilde{f}(\pi)\subset I(\tilde{f})\subset
Z(\tilde{f}_*(\pi),\pi)$, so $\tilde{f}_*(\pi)$ is abelian.
Hence $\tilde{f}_*$ is eventually commutative and $N(f)= R(f)
=R(\tilde{f}_*)=R(f_{1*}).$
The result now follows from theorem1.
\begin{example}
Let $f:X\rightarrow X$
 be a hyperbolic endomorphism of torus $T^k$.Then
$H_1(X,\bbbz)$ is torsion free and
\begin{equation}
N(f)= R(f) = (-1)^{r+p} \sum_{i=0}^k (-1)^i \tr (\Lambda^i {f_{1*}}).
\end{equation}
\end{example}

\subsection{Polyhedra with finite fundamental group.}

For a compact polyhedron $X$ with finite fundamental group,
$\pi_1(X)$, the universal cover $\tilde{X}$ is compact,
so we may explore the relation between $L(\tilde{f})$
and $\ind(p(Fix\ \tilde{f}))$.

\begin{definition}
The number $\mu([\tilde{f}])=\#Fix\ \tilde{f}_*$, defined to
be the order of the finite group $Fix\ \tilde{f}_*$, is called
the {\it multiplicity} of the lifting class $[\tilde{f}]$,
or of the fixed point class $p(Fix\ \tilde{f})$.
\end{definition}

\begin{lemma}[\cite{j}]
$$
L(\tilde{f}) = \mu([\tilde{f}]).\ind(f,p(Fix\ \tilde{f})).
$$
\end{lemma}
Let $W$ be the complex vector space of complex valued class functions on the
fundamental group $\pi$.The map $\tilde{f}_*$ induces a linear map
$B:W\rightarrow W$ defined by
$$
B(f):=f\circ\tilde{f}_*.
$$

\begin{theorem}
Let $X$ be a connected, compact polyhedron with finite fundamental group $\pi$.
Suppose that the action of $\pi$ on the rational homology of the
universal cover $\tilde{X}$ is trival,
i.e. for every covering translation $\alpha\in\pi$,
$\alpha_*=id:H_*(\tilde{X},\bbbq)\rightarrow H_*(\tilde{X},\bbbq)$.
Let $L(f)\not=0$.Then
\begin{equation}
N(f)=R(f)=TrB,
\end{equation}

\end{theorem}

{\sc Proof}
Under our assumption on $X$, any two liftings $\tilde{f}$
 and $\alpha\circ\tilde{f}$ induce the same homology homomorphism
 $H_*(\tilde{X},\bbbq)\rightarrow H_*(\tilde{X},\bbbq)$, and
 have thus the same value of $L(\tilde{f})$. From
 Lemma 8 it follows that any two fixed point classes $ f $ are either
 both essential or both inessential.
 Since $L(f)\not=0$ there is at least one
 essential fixed point class of $f$. Therefore all fixed point classes of $f$
 are essential and  $N(f)=R(f)$.The formula for $N(f)$ follows  now from
theorem 4

\begin{lemma}
Let $X$ be a polyhedron with finite fundamental group $\pi$ and let
$p:\tilde{X}\rightarrow X$ be its universal covering. Then the action
of $\pi$ on the rational homology of $\tilde{X}$ is trivial iff
$H_*(\tilde{X};\bbbq) \cong H_*(X;\bbbq)$.
\end{lemma}

\begin{corollary}
Let $\tilde{X}$ be a compact $1$-connected polyhedron which is a

 rational homology $n$-sphere, where $n$ is odd.
Let $\pi$ be a finite group acting freely on $\tilde{X}$ and let
$X=\tilde{X}/\pi$.
Then theorem 10 applies.
\end{corollary}

{\sc Proof}
The projection $p:\tilde{X}\rightarrow X= \tilde{X}/\pi$ is a
universal covering space of $X$. For every $\alpha\in\pi$, the degree
of $\alpha:\tilde{X}\rightarrow\tilde{X}$ must be 1, because
$L(\alpha)=0$ ($\alpha$ has no fixed points). Hence
$\alpha_* = id: H_*(\tilde{X};\bbbq)\rightarrow H_*(\tilde{X};\bbbq)$.

\begin{corollary}
If $X$ is a closed 3-manifold with finite $\pi$, then theorem 10
applies.
\end{corollary}

{\sc Proof}
$\tilde{X}$ is an orientable, simply connected manifold, hence a
 homology 3-sphere. We apply corollary .

\begin{corollary}
Let $X=L(m,q_1,\ldots,q_r)$ be a generalized lens space
and $f:X\to X$ a continuous map  with
$f_{1*}(1)=d$ where $ d \not= 1$.
Then theorem 10 applies.

\end{corollary}

{\sc Proof}
By corollary 1 we see that theorem 10
applies for lens spaces.
Since $\pi_1(X)=\bbbz/m\bbbz$,
the map $f$ is eventually commutative.
A lens space has a structure as a CW complex with one
cell $e_i$ in each dimension $0\leq i\leq 2l+1$.
The boundary map is given by $\partial e_{2k}=m.e_{2k-1}$
for even cells, and $\partial e_{2k+1}=0$ for odd cells.
{}From this we may calculate the Lefschetz numbers:
$ L(f) = 1-d^{(l+1)} \not= 0.$

\subsection{Other special cases}

\subsubsection{Self-map of a nilmanifold}
Theorem 6 implies
\begin{theorem}
Let $f$ be any continious map of a nilmanifold $M$ to itself.If $R(f)$ is
finite
then
\begin{equation}
N(f)= R(f) = (-1)^{r+p} \sum_{i=0}^m (-1)^i \tr \Lambda^i\tilde F,
\end{equation}
where $\tilde F$, $m$,$r$, and $p$ are the same as in theorem 6
\end{theorem}

 \subsubsection{ Pseudo-Anosov homeomorphism of a compact surface}
Let $X$ be a compact surface of negative euler characteristic and
 $f:X\rightarrow X$ is a pseudo-Anosov homeomorphism,i.e. there is a number
$\lambda >1$ and a pair of transverse measured foliations $(F^s,\mu^s)$ and
$(F^u,\mu^u)$ such that $f(F^s,\mu^s)=(F^s,\frac{1}{\lambda}\mu^s)$ and
$f(F^u,\mu^u)=(F^u,\lambda\mu^u)$.
Fathi and Shub \cite{fash} has proved the existence of Markov partitions for a
pseudo-Anosov homeomorphism.The existence of Markov partitions implies that
there is a symbolic dynamics for $(X,f)$.This means that there is a finite set
$N$, a matrix $A=(a_{ij})_{(i,j)\in N\times N}$ with entries $0$ or $1$ and a
surjective map $p:\Omega\rightarrow X$,where
$$
\Omega=\{(x_n)_{n\in \bbbz}: a_{x_nx_{n+1}}=1  ,  n\in \bbbz \}
$$
such that $p\circ \sigma =f\circ p$ where $\sigma$ is the shift (to the left)
of the sequence $(x_n)$ of symbols.We have first \cite{bl}:
$$
\# Fix\sigma ^n=\tr A^n.
$$
In general $p$ is not bijective.The non-injectivity of $p$ is due to the fact
that the rectangles of the Markov partition can meet on their boundaries.To
cansel the overcounting of periodic points on these boundaries,we use Manning's
combinatorial arguments \cite{m} proposed in the case of Axiom A diffeomorphism
(see also \cite {pp}) .Namely, we construct finitely many subshifts of finite
type ${\sigma _i}, i=0,1,..,m$, such that $\sigma_0=\sigma$, the other shifts
semi-conjugate with restrictions of $f$ \cite {pp} ,and signs $ \epsilon _i\in
\{-1,1\}$ such that for each $n$
$$
\# Fixf^n=\sum_{i=0}^m \epsilon_i.\#Fix \sigma_i^n =\sum_{i=0}^m \epsilon_i.\tr
A_i^n,
$$
where $A_i$ is transition matrix, corresponding to subshift of finite type
$\sigma_i$.
For pseudo-Anosov homeomorphism of compact surface  $N(f^n)=\# Fix(f^n)$ for
each $n>o$ \cite{th}.So we have following trace formula for Nielsen numbers

\begin{theorem}
Let $X$ be a compact surface of negative euler characteristic and
$f:X\rightarrow X$ is a pseudo-Anosov homeomorphism.Then
$$
N(f^n)=\sum_{i=0}^m \epsilon_i.\tr A_i^n.
$$
\end{theorem}

\subsubsection{Homeomorphisms of a hyperbolic 3-manifolds}

\begin{theorem}{\cite{jw}}
 Suppose $M$ is a orientable compact connected 3-manifold such that int$M$
admits a complete hyperbolic structure with finite volume and $ f: M
\rightarrow M $ is orientation preserving homeomorphism.Then
 $$
 N(f)=L(f)= \sum_{k=0}^{\dim M} (-1)^k \tr\Big[f_{*k}:H_k(X;\bbbq)\to
H_k(X;\bbbq)\Big]
$$

\end{theorem}

  \section[Reidemeister trace formula]{The Reidemeister trace formula for
generalised Lefschetz numbers}
 The results of this section are well known(see
\cite{r},\cite{w},\cite{fahu},\cite{j1}).We shall use this results later in
section  to estimate the radius of convergence of the Nielsen zeta function.
 The fundamental group  $\pi=\pi_1(X,x_0)$ splits into $\tilde{f}_*$-conjugacy
classes.Let $\pi_f$ denote the set of $\tilde{f}_*$-conjugacy classes,and
$\bbbz\ pi_f$ denote the abelian group freely generated by $\pi_f$ .We will use
the bracket notation $a\to [a]$ for both projections $\pi\to \pi_f$ and
$\bbbz\pi\to \bbbz\pi_f$.
Let $x$ be a fixed point of $f$.Take a path $c$ from $x_0$ to $x$.The
$\tilde{f}_*$-conjugacy class in $\pi$ of the loop $c.(f\circ c)^{-1}$,which is
evidently independent of the choice of $c$, is called the coordinate of $x$.Two
fixed points are in the same fixed point class $F$ iff they have the same
coordinates.This
$\tilde{f}_*$-conjugacy class is thus called the coordinate of the fixed point
class $F$ and denoted $cd_{\pi}(F,f)$ (compare with  discription in section 2).

The generalised Lefschetz number  or the Reidemeister trace \cite{r} is defined
as
\begin{equation}
L_{\pi}(f):=\sum_{F}ind(F,f).cd_{\pi}(F,f)  \in  \bbbz\pi_f ,
\end{equation}

the summation being over all essential fixed point classes $F$of $f$.The
Nielsen number $N(f)$ is the number of non-zero terms in $L_{\pi}(f)$,and the
indices of the essential fixed point classes appear as the coefficients in
$L_{\pi}(f)$.This invariant used to be called the Reidemeister trace because it
can be computed as an alternating sum of traces on the chain level  as follows
(\cite{r},\cite {w}) .
Assume that $X$ is a finite cell complex and  $f:X\to X$ is a cellular map.A
cellular decomposition ${e_j^d}$ of $X$ lifts to a $\pi$-invariant cellular
structure on the universal covering $\tilde X$.Choose an arbitrary lift
${\tilde{e}_j^d}$ for each ${e_j^d}$ .They constitute a free  $\bbbz\pi$-basis
for the cellular chain complex of $\tilde{X}$.The lift $\tilde{f}$ of $f$ is
also a cellular map.In every dimension $d$, the cellular chain map $\tilde{f}$
gives rise to a $\bbbz\pi$-matrix $ \tilde{F}_d $ with respect to the above
basis,i.e $\tilde{F}_d=(a_{ij})$ if $
\tilde{f}(\tilde{e}_i^d)=\sum_{j}a_{ij}\tilde{e}_j^d $,where $ a_{ij}\in
\bbbz\pi $.Then we have the Reidemeister trace formula

\begin{equation}
L_{\pi}(f)=\sum_{d}(-1)^d[Tr\tilde{F}_d] \in \bbbz\pi_f .
\end{equation}

\subsection{The mapping torus approach to the Reidemeister trace formula.}

Now we describe alternative approach to the Reidemeister trace formula proposed
recently by Jiang \cite{j1}. This approach is useful when we study the periodic
points of $f$,i.e. the fixed points of the iterates of $f$.

 The mapping torus $T_f$ of $f:X\rightarrow X$ is the space obtained from
$X\times [o,\infty )$ by identifiing $(x,s+1)$ with $(f(x),s)$ for all $x\in
X,s\in [0 ,\infty )$.On $T_f$ there is a natural semi-flow $\phi :T_f\times
[0,\infty )\rightarrow T_f, \phi_t(x,s)=(x,s+t)$ for all $t\geq 0$.Then the map
 $f:X\rightarrow X$ is the return map of the semi-flow $\phi $.A point $x\in X$
and a positive number $\tau >0$ determine the orbit curve $\phi _{(x,\tau
)}:={\phi_t(x)}_{0\leq t \leq \tau}$ in $T_f$.

Take the base point $x_0$ of $X$ as the base point of $T_f$.It is known that
the fundamental group $H:=\pi_ 1(T_f,x_0)$ is obtained from $\pi $ by adding a
new generator $z$ and adding the relations $z^{-1}gz=\tilde f_*(g)$ for all
$g\in \pi =\pi _1(X,x_0)$.Let  $H _c$ denote the set of conjugacy classes in $H
$. Let $\bbbz H $ be the integral group ring of $H $, and let $\bbbz H_c $ be
the free abelian group with basis $ H _c $.We again use the bracket notation $
a\rightarrow [a] $  for both projections $H \rightarrow H _c $ and $ \bbbz H
\rightarrow \bbbz H _c $.  If  $F^n$ is a fixed point class  of $f^n$, then
$f(F^n)$ is also fixed point class of $f^n$ and
 $ind(f(F^n),f^n)=ind(F^n,f^n)$.
Thus $f$ acts as an index-preserving permutation among fixed point classes of
$f^n$.By definition, an $n$-orbit class $O^n$  of $f$ to be the union of
elements of an orbit of this action.In other words, two points $x,x'\in
Fix(f^n)$ are said to be in the same $n$-orbit class of $f$ if and only if some
$f^i(x)$ and some $f^j(x')$ are in the same fixed point class of $f^n$.The set
$Fix(f^n)$ splits into a disjoint union of $n$-orbits classes.Point $x$ is a
fixed point of $f^n$
or a periodic point of period $n$ if and only if orbit curve  $\phi _{(x,n)}$
is a closed curve. The free homotopy class of the closed curve $\phi _{(x,n)}$
will be called the $H$ -coordinate of point $x$,written $cd_{H }(x,n)=[\phi
_{(x,n)}]\in H _c$.It follows that periodic points $x$ of period $n$ and $x'$
of period $n'$ have the same $H $-coordinate if and only if $n=n'$ and $x$,$x'$
belong to the same $n$-orbits class of $f$. Thus it is possible equivalently
define $x,x'\in Fixf^n $ to be in the same $n$-orbit class if and only if

Recently, Jiang \cite{j1} has considered generalised Lefschetz number with
respect to $H $
\begin{equation}
L_{H }(f^n):= \sum_{O^n}ind(O^n,f^n).cd_{H }(O^n) \in \bbbz H _c,
\end{equation}
and proved following trace formula:
\begin{equation}
L_{H }(f^n)=\sum_{d}(-1)^d[Tr(z\tilde{F}_d)^n] \in \bbbz H_c,
\end{equation}
where $\tilde{F}_d$ be $\bbbz \pi$-matrices defined in (16) and $z\tilde{F}_d$
is regarded as a $\bbbz H $-matrix.

\section[Group endomorphisms]{The Reidemeister zeta function of a group
endomorphism}

\begin{quotation}
 {\sc PROBLEM}.
For which groups and endomorphisms is the Reidemeister zeta function
a rational function? When does it have a functional equation?
Is $R_\phi(z)$ an algebraic function?
\end{quotation}

\subsection{ Reidemeister zeta functions
 of eventually commutative endomorphisms.}

As we remarked in section 1 to find out about the Reidemeister zeta functions
of eventually commutative endomorphisms, it is sufficient to
study the zeta functions of endomorphisms of abelian groups.
For the rest of this section G will be a finitely generated
abelian group.

\begin{theorem}
Let $G$ is a finitely generated abelian group and $\phi$ an
endomorphism of $G$ .Then $R_\phi(z)$ is a rational function and is equal to
\begin{equation}
 R_\phi(z)
 =
 \left(
 \prod_{i=0}^k
 \det(1-\Lambda^i\phi^\infty\otimes A .\sigma.z)^{(-1)^{i+1}}
 \right)^{(-1)^r}
\end{equation}
where matrix $A$ is defined in lemma 3, $\sigma=(-1)^p$,$p$ , $r$ and $k$ are
constants described in theorem 1.
\end{theorem}

{\sc Proof}
If we repeat the proof of the theorem 1 for $\phi^n$ instead $\phi$ we receive
that $ R(\phi^n)= R((\phi^\infty)^n.R((\phi^{tor})^n)$.From this and lemmas 2
and 3   we have the trace formula for $ R(\phi^n)$:
\begin{eqnarray*}
 R(\phi^n) & = & (-1)^{r+pn} \sum_{i=0}^k (-1)^i \tr \Lambda^i(\phi^\infty)^n .
\tr A^n \\
           & = &  (-1)^{r+pn} \sum_{i=0}^k (-1)^i \tr
(\Lambda^i(\phi^\infty)^n\otimes A^n) \\
         & = & (-1)^{r+pn} \sum_{i=0}^k (-1)^i \tr (\Lambda^i\phi^\infty\otimes
A )^n.
\end{eqnarray*}

We now calculate directly
\begin{eqnarray*}
R_\phi(z) & = & \exp\left(\sum_{n=1}^\infty \frac{R(\phi^n)}{n} z^n\right) \\
          & = & \exp\left(\sum_{n=1}^\infty
                \frac{(-1)^{r}
                \sum_{i=0}^k (-1)^i
                \tr (\Lambda^i\phi^\infty\otimes A )^n}{n} (\sigma .z)^n\right)
\\
          & = & \left(\prod_{i=0}^k\left(
 \exp\left(\sum_{n=1}^\infty
                \frac{1}{n}\tr (\Lambda^i\phi^\infty\otimes A )^n.(\sigma
.z)^n\right)
                \right)^{(-1)^i} \right)^{(-1)^r} \\
          & = & \left(\prod_{i=0}^k
                \det\left(1-\Lambda^i\phi^\infty\otimes A  .\sigma
.z\right)^{(-1)^{i+1}}
                \right)^{(-1)^r}.
\end{eqnarray*}

\subsection{Endomorphisms of finite groups}

\begin{theorem}
Let $\phi$ be an endomorphism of a finite group $G$.
Then $R_\phi(z)$ is a rational function and given by formula

 \begin{equation}
 R_\phi(z) = \frac{1}{\det(1-Bz)},
\end{equation}
Where $B$ is defined in theorem 4
\end{theorem}

{\sc Proof}
{}From theorem 4 it follows that $ R(\phi^n)=\tr B^n $ for every $n>o$.We now
calculate directly
\begin{eqnarray*}
R_\phi(z) & = &  \exp\left(\sum_{n=1}^\infty \frac{R(\phi^n)}{n} z^n\right)=
                      \exp\left(\sum_{n=1}^\infty \frac{\tr B^n}{n} z^n\right)=
 \exp\left(\tr \sum_{n=1}^\infty  \frac{ B^n}{n} z^n\right)\\
          &  = &  \exp\left(\tr (-log(1-Bz))\right)=  \frac{1}{\det(1-Bz)}.
\end{eqnarray*}

\subsection{Endomorphisms of the direct sum of a free abelian and a finite
group}

\begin{theorem}
Let $G$ is the direct sum of free abelian  and a finite group and $\phi$ an
endomorphism of $G$ .If the numbers $R(\phi^n)$ are all finite then $R_\phi(z)$
is a rational function and is equal to
\begin{equation}
  R_\phi(z)
 =
 \left(
 \prod_{i=0}^k
 \det(1-\Lambda^i\phi^\infty\otimes B .\sigma.z)^{(-1)^{i+1}}
 \right)^{(-1)^r}
\end{equation}
where matrix $B$ is defined in theorem 4, $\sigma=(-1)^p$,$p$ , $r$ and $k$ are
constants described in theorem 5.
\end{theorem}

{\sc Proof}
{}From lemma 4 it follows that $ R(\phi^n)=
R((\phi^\infty)^n.R((\phi^{tor})^n)$.
{}From now on the proof repeat the proof of the theorem 14.

\subsection{Endomorphisms of nilpotent groups}

\begin{theorem}
 If $G$ is a finitely generated torsion free nilpotent group and $\phi$ an
endomorphism of $G$ .Then $R_\phi(z)$ is a rational function and is equal to
\begin{equation}
 R_\phi(z)
 =
 \left(
 \prod_{i=0}^m
 \det(1-\Lambda^i\tilde F .\sigma.z)^{(-1)^{i+1}}
 \right)^{(-1)^r}
\end{equation}
where $\sigma=(-1)^p$,$p$ , $r$, $m$ and  $\tilde F$ is defined in section 1.5.
\end{theorem}

{\sc Proof}
If we repeat the proof of the theorem 6 for $\phi^n$ instead $\phi$ we receive
that $ R(\phi^n)=(-1)^{r+pn}\det(1-\tilde F)$( we suppose that Reidemeister
numbers  $ R(\phi^n)$ are finite for all $n$).Last formula implies the trace
formula for $ R(\phi^n)$ :

$$
R(\phi^n) = (-1)^{r+pn} \sum_{i=0}^m (-1)^i \tr (\Lambda^i\tilde F)^n
$$
{}From this we have  formula 22 immedeately by direct calculation as in theorem
14.

\begin{corollary}
Let the assumptions of theorem 17 hold.
Then the poles and zeros of the Reidemeister zeta function
are complex numbers which are reciprocal
of an eigenvalue of one of the matrices
 $$  \Lambda^i (\tilde F)
     : \Lambda^i (\tilde G)
       \longrightarrow
       \Lambda^i (\tilde G) \;\;\;\;\;\; 0\leq i\leq {\rm rank}\;G$$
\end{corollary}

\subsubsection{Functional equation}

\begin{theorem}
Let $\phi:G\to G$ be an endomorphism
of a finitely generated torsion free nilpotent group $G$.Then the Reidemeister
zeta function $R_\phi(z)$  has the
 following functional equation:
\begin{equation}
 R_{\phi}\left(\frac{1}{dz}\right)
 =
 \epsilon_2 . R_\phi(z)^{(-1)^{\rank G}}.
\end{equation}
where $d=\det\tilde F$ and $\epsilon_1$
are a constants in $\bbbc^\times$.
\end{theorem}

{\sc Proof}
Via the natural nonsingular pairing
$(\Lambda^i \tilde F)\otimes(\Lambda^{m-i} \tilde F) \rightarrow\bbbc$
the operators $\Lambda^{m-i}\tilde F$ and $d.(\Lambda^i \tilde F)^{-1}$ are
adjoint
to each other.

We consider an eigenvalue $\lambda$ of $\Lambda^i\tilde F$.
By theorem 17, This contributes a term
$$
\left((1-\frac{\lambda\sigma}{dz})^{(-1)^{i+1}}\right)^{(-1)^r}
$$
to $R_\phi\left(\frac{1}{dz}\right)$.

We rewrite this term as
 $$
 \left(
 \left( 1 - \frac{d\sigma z}{\lambda} \right)^{(-1)^{i+1}}
 \left( \frac{-dz}{\lambda\sigma} \right)^{(-1)^i}
 \right)^{(-1)^r}
 $$
 and note that $\frac{d}{\lambda}$ is an eigenvalue of $\Lambda^{m-i}\tilde F$.
Multiplying these terms together we obtain,
 $$
 R_\phi\left(\frac{1}{dz}\right)
 =
 \left(
 \prod_{i=1}^m
 \prod_{\lambda^{(i)}\in\spec\Lambda^i\tilde F}
 \left(\frac{1}{\lambda^{(i)}\sigma}\right)^{(-1)^i}
 \right)^{(-1)^r}
 \times
 R_\phi(z)^{(-1)^m}.
 $$
The variable $z$ has disappeared because
 $$
 \sum_{i=0}^m (-1)^i \dim\Lambda^i\tilde G = \sum_{i=0}^m (-1)^i  {C_k }^i = 0.
  $$

\subsection{Some conjectures for wider classes of groups}

For the case of almost nilpotent groups (ie. groups with polynomial growth, in
view of Gromov's theorem \cite{gromov}) we believe that some power of the
Reidemeister
 zeta function is a rational function.
We intend to prove this conjecture by
 identifying the Reidemeister number on the nilpotent part of the group
 with the number of fixed points in
 the direct sum of the duals of the quotients of successive terms in the
central series. We then hope to show that the Reidemeister number of the whole
endomorphism is a sum of numbers of orbits of such fixed points under the
 action of the finite quotient group (ie the quotient of the whole group by
 the nilpotent part).
The situation for groups with exponential growth is very different.
There one can expect the Reidemeister number to be infinite as long as the
endomorphism is injective. This can be proved in the case of
 surface  groups due to a theorem of C.Epstein (see \cite{epstein}).
He proves an estimate on numbers of geodesics, which when applied to the
 mapping torus of pseudo-Anosov map  guarantees infinitely
 many loops which wrap around the mapping torus exactly once.
This is equivalent (see section 4.1) to saying that the Reidemeister number is
infinite. A rigid hyperbolic group has a finite outer automorphism
group\cite{gromov1}.This implies that the Reidemeister number of some iteration
of endomorphism equals to the number of usual conjugacy classes which is in
this case infinite.
We believe this conjecture for the following reason.
Let $G$ be a group with exponential growth and let $l(g)$ be the length of an
element of $G$. Then one might expect that most of the time
$$
  l(g x \phi(g)^{-1}) > (1+\epsilon) l(g)
$$
for $g,x\in G$.
This would imply that for fixed $x$,
$$
 \#\{g x \phi(g)^{-1} : l(g x \phi(g)^{-1}) < N\}
 <
 \#\{g \in G : l(g x \phi(g)^{-1}) < N/(1+\epsilon)\}
$$
However since the group has exponential growth one can show
:
that this would imply
$$
 \frac{\#\{g x \phi(g)^{-1} : l(g x \phi(g)^{-1}) < N\}}
 {\#\{g \in G : l(g x \phi(g)^{-1}) < N\}}
 \to 0 \hbox{ as }N\to \infty.
$$
If there were only finitely many twisted conjugacy classes then
 we could derive a contradiction by summing the left hand side of the above
 formula over a set of representatives $x$ for the classes,
 and observing that this sum is always equal to 1.


\section[Continuous maps]{The Reidemeister and Nielsen zeta functions   of a
continuous map.}

\begin{remark}
Using lemma 6 we may apply all theorems of section 5 to the Reidemeister zeta
functions of continuos maps.
\end{remark}

\subsection{The Jiang subgroup and the Nielsen zeta function}

\begin{theorem}
Suppose that there is an integer $m$ such that $\tilde{f}_*^m(\pi)\subset
I(\tilde{f}^m)$.If $L(f^n)\not=0$ for every $n>o$ ,then
\begin{equation}
  N_f(z)=R_f(z)= \left(
 \prod_{i=0}^k
 \det(1-\Lambda^if_{1*}^\infty\otimes A .\sigma.z)^{(-1)^{i+1}}
 \right)^{(-1)^r}
\end{equation}
If  $L(f^n)=0$ only for finite number of $n$,then
\begin{equation}
 N_f(z)= \exp\left(P(z)\right).R_f(z)= \exp\left(P(z)\right).
 \left(
 \prod_{i=0}^k
 \det(1-\Lambda^if_{1*}^\infty\otimes A .\sigma.z)^{(-1)^{i+1}}
 \right)^{(-1)^r}
\end{equation}
Where $P(z)$ is a polynomial ,$A, k, p,$ and $r$ are as in theorem 14.

 \end{theorem}

{\sc Proof}

If $L(f^n)\not=0$ for every $n>o$ ,then formula (19) follows from theorems 9
and 14.
If  $L(f^n)=0$, then $N(f^n)=0$ .If $L(f^n)\not=0$, then
 $N(f^n)=R(f^n)$(see proof of theorem 9 ).So the fraction
$N_f(z)/R_f(z)=\exp(P(z))$, where $P(z)$ is a polynomial whose degree equal to
maximal $n$, such that  $L(f^n)\not=0$.

\begin{corollary}
Let the assumptions of theorem 19 hold.
Then the poles and zeros of the Nielsen zeta function
are complex numbers which are the reciprocal
of an eigenvalue of one of the matrices
 $$  \Lambda^i (f_{1*}^\infty\otimes A .\sigma) $$
\end{corollary}

\begin{corollary}
Let $I(id_{\tilde{X}})=\pi$ .If the assumptions of theorem 19 about Lefschetz
numbers hold,then formulas (24) and (25) are  valid.
\end{corollary}

\begin{corollary}
Suppose that $X$ is asphericaland $f$ is eventually commutative.If the
assumptions of theorem 19 about Lefschetz numbers hold,then formulas (24) and
(25) are  valid.
\end{corollary}

\subsection{Polyhedra with finite fundamental group and Nielsen zeta function}

\begin{theorem}
Let $X$ be a connected, compact polyhedron with finite fundamental group $\pi$.
Suppose that the action of $\pi$ on the rational homology of the
universal cover $\tilde{X}$ is trival,
i.e. for every covering translation $\alpha\in\pi$,
$\alpha_*=id:H_*(\tilde{X},\bbbq)\rightarrow H_*(\tilde{X},\bbbq)$.
If $L(f^n)\not=0$ for every $n>o$ ,then
\begin{equation}
  N_f(z)=R_f(z) = \frac{1}{\det(1-Bz)},
\end{equation}
If  $L(f^n)=0$ only for finite number of $n$,then
\begin{equation}
  N_f(z)=
\exp\left(P(z)\right).R_f(z)=\exp\left(P(z)\right).\frac{1}{\det(1-Bz)},
\end{equation}
Where $P(z)$ is a polynomial,   $B$ is defined in theorem 4
\end{theorem}

{\sc Proof}

If $L(f^n)\not=0$ for every $n>o$ ,then formula (26) follows from theorems 10
and 15.
If  $L(f^n)=0$, then $N(f^n)=0$ .If $L(f^n)\not=0$, then
 $N(f^n)=R(f^n)$(see proof of theorem 10).So the fraction
$N_f(z)/R_f(z)=\exp(P(z))$, where $P(z)$ is a polynomial whose degree equal to
maximal $n$, such that  $L(f^n)\not=0$.

\begin{corollary}
Let $\tilde{X}$ be a compact $1$-connected polyhedron which is a
 rational homology $n$-sphere, where $n$ is odd.
Let $\pi$ be a finite group acting freely on $\tilde{X}$ and let
$X=\tilde{X}/\pi$.
Then theorem 20 applies.
\end{corollary}

\begin{corollary}
 If $X$ is a closed 3-manifold with finite $\pi$, then theorem 20
applies.
\end{corollary}

\begin{example}[\cite{bb}]

Let $f:S^2\vee S^4\rightarrow S^2\vee S^4$ to be a continuous map of the
bouquet of spheres such that the restriction $f/_{S^4}=id_{S^4}$ and the degree
of the restriction $f/_{S^2}:S^2\rightarrow S^2$ equal to $-2$.Then $L(f)=0$,
hence
$N(f)=0$ since $ S^2\vee S^4$ is simply connected.For $k>1$ we have
$L(f^k)=2+(-2)^k\not=0$,therefore $N(f^k)=1$.From this we have by direct
calculation that
\begin{equation}
N_f(z)=\exp(-z).\frac{1}{1-z} .
\end{equation}
\end{example}

\begin{remark}

We would like to mention that in all known cases the Nielsen zeta function is a
nice function. By this we mean that it is a product of an exponential of a
polynomial with a function some power of which is rational. May be this is a
general pattern; it could however be argued that this just reflects our
inability to calculate the Nielsen numbers in general case.
\end{remark}

\subsection{Nielsen zeta function in other special cases}

Theorem 17 implies
\begin{theorem}
Let $f$ be any continious map of a nilmanifold $M$ to itself.If $R(f^n)$ is
finite for all $n$
then
\begin{equation}
N_f(z)= R_f(z)
 =
  \left(
 \prod_{i=0}^m
 \det(1-\Lambda^i\tilde F .\sigma.z)^{(-1)^{i+1}}
 \right)^{(-1)^r}
\end{equation}
where $\sigma=(-1)^p$,$p$ , $r$, $m$ and  $\tilde F$ is defined in section 5.4.
\end{theorem}

Theorem 12 implies

\begin{theorem}
Let $X$ be a compact surface of negative euler characteristic and
$f:X\rightarrow X$ is a pseudo-Anosov homeomorphism.Then
\begin{equation}
N_f(z)=\prod_{i=0}^m\det(1-A_iz)^{-\epsilon_i}
\end{equation}
where $A_i$ and $\epsilon_i$
 the same as in theorem 12.
\end{theorem}

Theorem 13 implies

\begin{theorem}
 Suppose $M$ is a orientable compact connected 3-manifold such that int$M$
admits a complete hyperbolic structure with finite volume and $ f: M
\rightarrow M $ is orientation preserving homeomorphism.Then Nielsen zeta
function is rational and
 $$
N_f(z)=L_f(z)
$$

\end{theorem}

\subsection{Radius of convergence of the Nielsen zeta function}

We denote by $R$ the radius of convergence of the Nielsen zeta function
$N_f(z)$, and by $h(f)$ the topological entropy of continuous map $f$. Let
$h=\inf  \{h(g)$ over all maps $g$ of the same homotopy type as $f$\}.
\begin {theorem}[\cite{fp}]
For any continuous map $f$ of any compact polyhedron $X$ into itself the
Nielsen zeta function has positive radius of convergence $R$ and
\begin{equation}
R \geq \exp(-h) > 0.
\end{equation}

\end{theorem}

In this section we propose another prove of positivity of $R$ and give an exact
 algebraic lower estimation for the radius $R$
using trace formulas (16) and (18)  for generalised Lefschetz numbers.

 For any set $S$ let $\bbbz S$ denote the free abelian group with the specified
basis $S$.The norm in $\bbbz S$ is defined by
\begin{equation}
\|\sum_i k_is_i\|:= \sum_i \mid k_i\mid \in \bbbz,
\end{equation}
when the $s_i$ in $S$ are all different.

For a $\bbbz H $-matrix $ A=(a_{ij}) $,define its norm by $ \| A
\|:=\sum_{i,j}\| a_{ij} \| $.Then we have inequalities $\| AB \|\leq \| A\ |.\|
B \|$ when $A,B$ can be multiplied, and $\| \tr A \|\leq \| A \| $ when $A$ is
a square matrix.For a matrix  $ A=(a_{ij})N(f^n)             $ in $\bbbz S$,
its matrix of norms is defined to be the matrix $ A^{norm}:=(\| a_{ij} \|)$
which is a matrix of non-negative integers.In what follows, the set $S$ wiill
be $\pi $, $H $ or
$H _c$.We denote by $s(A)$ the spectral radius of $A$, $s(A)=\lim_n (\| A^n \|
|)^{\frac{1}{n}},$ which coincide with the largest  modul of an eigenvalue of
$A$.

\begin{theorem}
For any continious map $f$ of any compact polyhedron   $X$ into itself the
Nielsen zeta function has positive radius of convergence $R$,which admits
following estimations
\begin{equation}
 R\geq \frac{1}{\max_d \| z\tilde F_d \|} > 0
\end{equation}
and
\begin{equation}
R\geq \frac{1}{\max_d s(\tilde F_d^{norm})} > 0,
\end{equation}
where $\tilde F_d$ is the same as in section 4.

\end{theorem}

{\sc Proof}
By the homotopy type invariance of the invariants we can suppose that $f$ is a
cell map of a finite cell complex.By the definition the Nielsen number $N(f^n)$
is the number of non-zero terms in $L_{\pi}(f^n)$(see section 4).The norm $\|
L_{H}(f^n) \| $ is the sum of absolute values of the indices of all the
$n$-orbits classes $O^n$ .It equals $\| L_{\pi}(f^n) \|$, the sum of absolute
values of the indices of all the fixed point classes of $f^n$, because any two
fixed point classes of $f^n$ contained in the same $n$-orbit class $O^n$ must
have the same index.From this we have
$N(f^n)\leq\|L_{\pi}(f^n)\|=\|L_H(f^n)\|=\|\sum_d(-1)^d[\tr(z\tilde
F_d)^n]\|\leq\sum_d\|[\tr(z\tilde F_d)^n]\|\leq\sum_d\|\tr(z\tilde
F_d)^n\|\leq\sum_d\|(z\tilde F_d)^n\|\leq\sum_d\|(z\tilde F_d)\|^n $(see
\cite{j1}).The radius of convergence $ R$ is given by Caushy-Adamar formula:
$$
 \frac{1}{R}=\limsup_n (\frac{N(f^n)}{n})^{\frac{1}{n}}=\limsup_n
(N(f^n))^{\frac{1}{n}}.
$$
Therefore we have:
$$
R= \frac{1}{\limsup_n (N(f^n))^{\frac{1}{n}}}\geq \frac{1}{\max_d \| z\tilde
F_d \|} > 0.
$$
Inequalities:
$$
N(f^n)\leq\|L_{\pi}(f^n)\|\!=\!\|L_H(f^n)\|\!=\!\|\sum_d(-1)^d[\tr(z\tilde
F_d)^n]\|\!\leq\sum_d\| [\tr(z\tilde F_d)^n]\|\!\leq
$$
$$
\sum_d{||}\tr(z\tilde F_d)^n{||}\leq \sum_d\tr((z\tilde
F_d)^n)^{norm}\leq\sum_d\tr((z\tilde F_d)^{norm})^n
\leq \sum_d\tr((\tilde F_d)^{norm})^n
$$
and the definition of spectral radius give estimation:
$$
R= \frac{1}{\limsup_n (N(f^n))^{\frac{1}{n}}} \geq \frac{1}{\max_d s(\tilde
F_d^{norm})} > 0.
$$

\begin{example}
 Let $X$ be surface with boundary, and  $f:X\rightarrow X$ be a map.Fadell and
Husseini \cite{fahu} devised a method of computing the matrices of the lifted
chain map for surface maps.Suppose $ \{a_1, .... ,a_r\} $ is a free basis for
$\pi_1(X)$. Then $X$ has the homotopy type of a bouquet $B$  of $r$ circles
which can be decomposed into one 0-cell and $r$  1-cells corresponding to the
$a_i$,and $f$ has the homotopy type of a cellular map  $g:B\rightarrow B.$
By the homotopy type invariance of the invariants,we can replace $f$ with $g$
in computations.The homomorphism $\tilde f_*:\pi_1(X)\rightarrow \pi_1(X)$
induced by $f$ and $g$ is determined by the images $b_i=\tilde f_*(a_i), i=1,..
,r $.The fundamental group $\pi_1(T_f)$ has a presentation
$\pi_1(T_f)=<a_1,...,a_r,z| a_iz=zb_i, i=1,..,r>$.Let
$$
D=(\frac{\partial b_i}{\partial a_j})
$$
be the Jacobian in Fox calculus(see \cite{bi}).Then,as pointed out in
\cite{fahu}, the matrices of the lifted chain map $\tilde g$ are
$$
\tilde F_0=(1), \tilde F_1=D=(\frac{\partial b_i}{\partial a_j}).
$$
Now, we can find estimations for the radius $R$ from (33) and (34).

\end{example}
\begin{remark}
 Let $X$ be a compact connected surface and  $f:X\rightarrow X$ be a
homeomorphism.Let $\chi(X)<0$.By Thurston's classification theorem \cite{th}
$f$ is isotopic to a homeomorphism $\phi$ such that either (1) $\phi$ is a
periodic map;or (2) $\phi$ is a pseudo-Anosov map with stretching factor
$\lambda > 1$(see section 3.3.1);or  (3) $\phi$ is a reducible map,i.e. there
is a system of disjoint simple closed curves in int$X$ which is invariant by
$\phi$ and which has a$\phi$-invariant tubular neighborhood $U$ such that each
component of $X-U$ has negative euler characteristic and on each
$\phi$-component of $X-U$ , $\phi$ satisfies \(1\) or \(2\).Then, as follows
from \cite{fp},radius $R=\frac{1}{\lambda}$, where $\lambda > 1$ is the largest
stretching factor of the pseudo-Anosov pieses(if there is no pseudo-Anosov
piece)   $\lambda=1$.

\end{remark}

\section{Connection with Reidemeister Torsion}

\subsection{Preliminaries}

\subsubsection{Reidemeister torsion}

Like the Euler characteristic, the Reidemeister torsion is algebraically
defined.
Roughly speaking, the Euler characteristic is a
graded version of the dimension, extending
the dimension from a single vector space to a complex
of vector spaces.
In a similar way, the Reidemeister torsion
is a graded version of the absolute value of the determinant
of an isomorphism of vector spaces.
Let $d^i:C^i\rightarrow C^{i+1}$ be a cochain complex $C^*$
of finite dimensional vector spaces over $\bbbc$ with
$C^i=0$ for $i<0$ and large $i$.
If the cohomology $H^i=0$ for all $i$ we say that
$C^*$ is {\it acyclic}.
If one is given positive densities $\Delta_i$ on $C^i$
then the Reidemeister torsion $\tau(C^*,\Delta_i)\in(0,\infty)$
for acyclic $C^*$ is defined as follows:

\begin{definition}
 Consider a chain contraction $\delta^i:C^i\rightarrow C^{i-1}$,
 ie. a linear map such that $d\circ\delta + \delta\circ d = id$.
 Then $d+\delta$ determies a map

 $ (d+\delta)_+ : C^+:=\oplus C^{2i}
                      \rightarrow C^- :=\oplus C^{2i+1}$
 and a map
 $ (d+\delta)_- : C^- \rightarrow C^+ $.
Since the map
$(d+\delta)^2 = id + \delta^2$ is unipotent,
$(d+\delta)_+$ must be an isomorphism.
One defines $\tau(C^*,\Delta_i):= \mid \det(d+\delta)_+\mid$
(see \cite{fri2}).
\end{definition}

Reidemeister torsion is defined in the following geometric setting.
Suppose $K$ is a finite complex and $E$ is a flat, finite dimensional,
complex vector bundle with base $K$.
We recall that a flat vector bundle over $K$ is essentially the
same thing as a representation of $\pi_1(K)$ when $K$ is
connected.
If $p\in K$ is a basepoint then one may move the fibre at $p$
in a locally constant way around a loop in $K$. This

defines an action of $\pi_1(K)$ on the fibre $E_p$ of $E$
above $p$. We call this action the holonomy representation
$\rho:\pi\to GL(E_p)$.
Conversely, given a representation $\rho:\pi\to GL(V)$
of $\pi$ on a finite dimensional complex vector space $V$,
one may define a bundle $E=E_\rho=(\tilde{K}\times V) / \pi$.
Here $\tilde{K}$ is the universal cover of $K$, and
$\pi$ acts on $\tilde{K}$ by covering tranformations and on $V$
by $\rho$.
The holonomy of $E_\rho$ is $\rho$, so the two constructions
give an equivalence of flat bundles and representations of $\pi$.

If $K$ is not connected then it is simpler to work with
flat bundles. One then defines the holonomy as a
representation of the direct sum of $\pi_1$ of the
components of $K$. In this way, the equivalence of
flat bundles and representations is recovered.

Suppose now that one has on each fibre of $E$ a positive density
which is locally constant on $K$.
In terms of $\rho_E$ this assumption just means
$\mid\det\rho_E\mid=1$.
Let $V$ denote the fibre of $E$.

Then the cochain complex $C^i(K;E)$ with coefficients in $E$
can be identified with the direct sum of copies
of $V$ associated to each $i$-cell $\sigma$ of $K$.
The identification is achieved by choosing a basepoint in each
component of $K$ and a basepoint from each $i$-cell.
By choosing a flat density on $E$ we obtain a
preferred density $\Delta_i$ on $C^i(K,E)$. One defines the
R-torsion of $(K,E)$ to be
$\tau(K;E)=\tau(C^*(K;E),\Delta_i)\in(0,\infty)$.

\subsubsection{Unitary dual of the direct sum of a free abelian and a finite
group}

In this section let again the group $G$ will be $G=\bbbz^m\oplus F$ ,where $F$
is a finite group and $m$ is a natural number.Let $\phi: G\rightarrow G$ be an
endomorphism.One defines the unitary dual $\hat {G}$ of $G$ to be the set of
all equivalence classes of irreducible, unitary representations of $G$. If
$\rho:G\rightarrow U(V)$ is a unitary representation of $G$ on $V$ then
$\rho\circ \phi:G\rightarrow U(V)$ is also a representation, which we denote
$\hat \phi(\rho)$. If the representations $\rho_1$ and $\rho_2$ are equivalent
then representations $\hat \phi(\rho_1)$ and $\hat \phi(\rho_2)$ are also
equivalent. Therefore the endomorphism $\phi $ induces a dual map $\hat
{\phi}:\hat{G}\rightarrow \hat{G}$ from the unitary dual to itself. Now
consider the dual $\hat{G}$ for  $G=\bbbz^r\oplus F$. This is the cartesian
product of the duals of $\bbbz^m$ and $\hat{F}$: $\hat{G}=\hat{\bbbz^m}\otimes
\hat{F}$. From this it follows that topologically the dual $\hat{G}$ is a uni!
 on of finitely many disjoint tori

\begin{lemma}\cite{hi}
If $\phi$ is any endomorphism of $G$ where $G$ is the direct sum of a finite
group with a free abelian group, then
$$
 R(\phi )=\# Fix(\hat {\phi})
$$
\end{lemma}

\subsection{The Reidemeister zeta Function and the Reidemeister
Torsion of the Mapping Tori of the dual map.}

Let $f:X\rightarrow X$ be a homeomorphism of
a compact polyhedron $X$.

Let $T_f := (X\times I)/(x,0)\sim(f(x),1)$ be the
mapping tori of $f$.
We shall consider the bundle $p:T_f\rightarrow S^1$
over the circle $S^1$.
We assume here that $E$ is a flat, complex vector bundle with
finite dimensional fibre and base $S^1$. We form its pullback $p^*E$
over $T_f$.
Note that the vector spaces $H^i(p^{-1}(b),c)$ with
$b\in S^1$ form a flat vector bundle over $S^1$,
which we denote $H^i F$. The integral lattice in
$H^i(p^{-1}(b),\bbbr)$ determines a flat density by the condition
that the covolume of the lattice is $1$.
We suppose that the bundle $E\otimes H^i F$ is acyclic for all
$i$. Under these conditions D. Fried \cite{fri2} has shown that the bundle
$p^* E$ is acyclic, and we have
\begin{equation}
 \tau(T_f;p^* E) = \prod_i \tau(S^1;E\otimes H^i F)^{(-1)^i}.
\end{equation}
Let $g$ be the prefered generator of the group
$\pi_1 (S^1)$ and let $A=\rho(g)$ where
$\rho:\pi_1 (S^1)\rightarrow GL(V)$.
Then the holonomy around $g$ of the bundle $E\otimes H^i F$
is $A\otimes f^*_i$.

Since $\tau(E)=\mid\det(I-A)\mid$ it follows from (20)
that
\begin{equation}
 \tau(T_f;p^* E) = \prod_i \mid\det(I-A\otimes f^*_i)\mid^{(-1)^i}.
\end{equation}
We now consider the special case in which $E$ is one-dimensional,
so $A$ is just a complex scalar $\lambda$ of modulus one.
Then in terms of the rational function $L_f(z)$ we have \cite{fri2}:
\begin{equation}
 \tau(T_f;p^* E) = \prod_i \mid\det(I-\lambda .f^*_i)\mid^{(-1)^i}
             = \mid L_f(\lambda)\mid^{-1}
\end{equation}

\begin{theorem}
Let $\phi:G\to G$ be an automorphism of $G$,where  $G$ is the direct sum of a
finite group with a finitely generated free abelian group, then
$$
 \tau\left(T_{\hat{\phi}};p^*E\right)
 =
 \mid L_{\hat{\phi}}(\lambda) \mid^{-1}
 =
 \mid R_{\phi}(\sigma\lambda) \mid^{(-1)^{r+1}},
$$
where $\lambda$ is the holonomy of the one-dimensional
flat complex bundle $E$ over $S^1$, $r$ and $\sigma$ are the constants
described in theorem 5 .
\end{theorem}

{\sc Proof}
We  know from lemma 10
that $R(\phi^n)$ is the number of fixed points
of the map $\hat{\phi}^{n}$.
In general it is only necessary to check that
the number of fixed points of $\hat{\phi}^n$ is equal to the
absolute value of its Lefschetz number.
We assume without loss of generality that $n=1$.
We are assuming that $R(\phi)$ is finite, so
the fixed points of $\hat{\phi}$ form a discrete set.
We therefore have
 $$
 L(\hat{\phi})
 =
 \sum_{x\in\fix\hat{\phi}} \ind(\hat{\phi},x).
 $$
Since $\phi$ is a group endomorphism, the trivial representation
$x_0\in\hat{G}$ is always fixed.
Let $x$ be any fixed point of $\hat{\phi} $.Since $\hat{G}$ is union of tori
$\hat{G}_0,...,\hat{G}_t$ and $\hat{\phi}$ is a linear map, we  can
shift any two fixed points onto one another without altering the map
$\hat\phi$.
This gives us for any fixed point $x$ the equality
 $$ \ind(\hat{\phi},x) = \ind(\hat{\phi},x_0) $$
and so all fixed points have the same index.
It is now sufficient to show that $\ind(\hat{\phi},x_0)=\pm 1$.
This follows because the map on the torus
 $$ \hat{\phi}:\hat{G}_0\to\hat{G}_0 $$
lifts to a linear map of the universal cover, which is an euclidean space. The
index is then the sign of the determinant of the identity map  minus this
lifted map.
This determinant cannot be zero, because $1-\hat{\phi}$ must have finite
kernel by our assumption that the Reidemeister number of $\phi$ is
finite
(if $\det(1-\hat{\phi})=0$ then the kernel of $1-\hat{\phi}$

is a positive dimensional subspace of $\hat{G}$, and therefore
infinite).

\begin{corollary}
Let $f:X\to X$ be a homeomorphism of a compact
polyhedron $X$. If $\pi_1(X)$ is the direct sum of a finite group with a free
abelian group, then
then
$$
 \tau\left(T_{\widehat{(f_{1*})}};p^*E\right)
 =
 \left|\ L_{\widehat{(f_{1*})}}(\lambda) \ \right|^{-1}
 =
 \Big|\ R_f(\sigma\lambda) \ \Big|^{(-1)^{r+1}},
$$
where $r$ and $\sigma$ are  the constants described in theorem 9.
\end{corollary}

\section{Concluding remarks}
\subsection{Reidemeister and Nielsen numbers and zeta functions  modulo
a normal subgroup}

In the theory of (ordinary) fixed point classes,
we work on the universal covering space.
The group of covering transformations plays a key role.
It is not surprising that this theory can be generalized
to work on all regular covering spaces.
Let $K$ be a normal subgroup of the fundamental group $\pi_1(X)$.
Consider the regular covering $p_K:\tilde{X}/K \to X$
corresponding to $K$.
A map $\tilde{f}_K:\tilde{X}/K \to\tilde{X}/K$
is called a lifting of $f:X\to X$ if
$p_K\circ\tilde{f}_K = f\circ p_K$.
We know from the theory of covering spaces that such liftings exist
if and only if $f_*(K)\subset K$.
If $K$ is a fully invariant subgroup of $\pi_1(X)$ ( in the
sense that every endomorphism sends $K$ into $K$)
such as, for example the commutator subgroup of $\pi_1(X)$,
then there is a lifting of any continuous map.
We may define the ${\mod K}$-Reidemeister and Nielsen numbers (see \cite{j})
and
zeta functions (see \cite{f3}) and develop a similar
theory for them by simply replacing $\tilde{X}$ and
$\pi_1(X)$ by $\tilde{X}/K$ and $\pi_1(X)/K$ in every definition,
every theorem and every proof, since everything was done in terms

of liftings and covering translations.

\subsection{Minimal dynamical zeta function}

\subsubsection{Radius of convergence of the minimal zeta function}

In the Nielsen theory for periodic points, it is well known that $N(f^n)$ is
often poor as a lower bound for the number of fixed points of $f^n$.A good
homotopy invariant lower bound $ NF_n(f)$,called the Nielsen type number for
$f^n$,is defined in \cite{j}.Consider any finite set of periodic orbit classes
$\{O^{k_j}\}$ of varied period $k_j$ such that every essential periodic
$m$-orbit class, $m|n$, contains at least one class in the set.Then $NF_n(f)$
is the minimal sum $ \sum_j k_j$ for all such finite sets.
Halpern(see \cite{j}) has proved that for all $n$ $ NF_n(f) =\min\{\#Fix(g)| g$
has the same homotope type as $f$ \}.Recently, Jiang \cite{j1} found that as
far as asymptotic growth rate is concerned,these Nielsen type numbers are no
better than the Nielsen numbers.
\begin{lemma}[\cite{j1}]
\begin{equation}
\limsup_n (N(f^n))^{\frac{1}{n}}=\limsup_n (NF_n(f))^{\frac{1}{n}}
\end{equation}
\end{lemma}
Let us consider minimal dynamical zeta function
$$
 M_f(z) := \exp\left(\sum_{n=1}^\infty \frac{NF_n(f)}{n} z^n \right),
$$
In paper \cite{f4} it was proved that $ M_f(z) $ has positive radius of
convergence $R$. Below we propose another prove of this fact and give an exast
algebraic
 lower estimation for the radius $R$.
\begin{theorem}
For any continious map $f$ of any compact polyhedron   $X$ into itself the
minimal zeta function has positive radius of convergence $R$,which admits
following estimations
\begin{equation}
R \geq \exp(-h) > 0,
\end{equation}
 \begin{equation}
R\geq \frac{1}{\max_d \| z\tilde F_d \|} > 0,
\end{equation}
and
\begin{equation}
R\geq \frac{1}{\max_d s(\tilde F_d^{norm})} > 0,
\end{equation}
where $\tilde F_d$ and $h$ is the same as in section 6.1.

\end{theorem}

{\sc Proof}
The theorem follows from Caushy-Adamar formula,lemma 11 and theorems 24 and 25.

\subsection{Dold congruenses}

\begin{definition}
We shall say that a sequence of integers $a(n)$ satisfies the
 {\sl Dold congruences}
 iff for all $ n\in N$ one has
$$
 \sum_{d\mid n} \mu(d) a(n/d) \equiv 0 \mod n.
$$
\end{definition}

Dold \cite{do} proved that the sequence $L(f^n)$ of Lefschetz
 numbers has this property.
The following easy lemma can be proved purely combinatorially:

\begin{lemma}
 Let $X$ is any set and $f:X\to X$ any function.
If for any $n$, $f^n$ has only finitely many fixed points in $X$ then
the sequence of numbers $\# Fix(f^n)$ satisfies the Dold congruences.
Any sequence of natural numbers satisfying the Dold congruences
  arises in this way.
\end{lemma}

In \cite{fh1} we have proved that if  $f:X\to X$ is eventually commutative map
of compact polyhedron,or any self-map if $\pi_1(X)$ is finite, then the numbers
$R(f^n)$ satisfy the Dold congruences.\\
 Let $g:M\to M$ be an expanding map \cite{sh} of the orientable smooth compact
manifold $M$.Then $M$ is aspherical and is a $K(\pi_1(M),1)$, and $\pi_1(M)$ is
torsion free \cite{sh}.According to Shub \cite{sh} any lifting $\tilde g$ of
$g$ has exactly one fixed point.From this and the covering homotopy theorem it
follows that the fixed point of $g$ are pairwise inequivalent.The same is true
for all iterates $g^n$.Therefore $N(g^n)=NF_n(g)=\# Fix(g^n)$ for all $n$.So
the sequence of the  Nielsen and Nielsen type numbers $N(g^n)=NF_n(g)$ of an
expanding map satisfies the Dold congruences.\\
For a hyperbolic endomorphism of torus $T^n$ and for a pseudo-Anosov
homeomorphism of a compact surface we also have equalities $N(g^n)=NF_n(g)=\#
Fix(g^n)$ for all $n$ (see \cite{f4} and \cite{th}), and ,therefore, the  Dold
congruences   for the sequence of the  Nielsen and Nielsen type numbers .

\subsection{Open questions}

\begin{question}
For which spases and maps in the inequalities (33) - (34) does the equality
hold
?
\end{question}

Sometime, the minimal zeta function coincide with the Nielsen zeta function,for
example, for hyperbolic endomorphism of torus, for an expanding map of the
orientable smoth manifold (see \cite{f4}).This motivate the following

\begin{question}
For which spaces and maps minimal zeta function is a rational function,when it
is a meromorphic function?When does it have a functional equation?Which zeros
and poles it has?
\end{question}

\begin{question}
For which spases and maps we have trace formula for Nielsen type numbers?
When we have  the Dold congruences  for this numbers?When  in the inequalities
(37)-(38) does equalities hold?
\end{question}

The trace formulae which we recieve in this article
 appear to be very similar to formulae arising in thermodynamical formalism.
The relation of the radius of convergence with the entropy (Theorem 24), Markov
partition in the pseudo-Anosov case(Theorem 22) and the relation of the
Reidemeister zeta function to Artin-Mazur zeta function on the unitary dual
space(lemma 10) also indicate  a connection with this theory. It is another
open question to understand this connection.

\bigskip

IHES, 35 route de Chartres, 91440- Bures sur Yvette, France .

{\it E-mail address}: felshtyn@ihes.fr

Ernst-Moritz-Arndt Universit\"at, Fachbereich Mathematik / Informatik,

Jahn-strasse 15a, D-17489 Greifswald, Germany.

{\it E-mail address}: felshtyn@math-inf.uni-greifswald.d400.de

\bigskip
Max-Planck-Institut f\"ur Mathematik, Gottfried-Claren-Strasse 26,

D-5300 Bonn 3, Germany.
{\it E-mail address}: hill@mpim-bonn.mpg.de

\end{document}